\documentclass[longauth]{aa}  

\usepackage{xcolor}
\usepackage[colorlinks=true, citecolor=blue, linkcolor=blue, urlcolor=blue]{hyperref}

\usepackage{natbib}
\usepackage[switch]{lineno} 

\usepackage{graphicx}
\usepackage{caption}
\usepackage{subcaption}
\usepackage{lipsum}
\usepackage{inputenc}
\usepackage{color}
\usepackage{amsmath,amsfonts,amssymb}
\usepackage{float}
\usepackage{hyperref}
\usepackage[version=3]{mhchem}
\usepackage{lineno}
\usepackage{amsmath} 
\usepackage{cuted}
\usepackage{lipsum}
\usepackage{mathtools}
\usepackage{xcolor}
\usepackage{multirow}
\usepackage{fleqn}
\usepackage{graphicx,animate} 
\usepackage{subcaption}
\usepackage{tabularx} 
\usepackage{booktabs} 
\usepackage{overpic}
\usepackage{rotating}

\newcommand{\vpm}[2]{\begin{tabular}[t]{@{}c@{}}$#1$ \\ {\fontsize{9}{10}\selectfont$(\pm\, #2)$}\end{tabular}}

\begin{document}

   \title{Energy distributions of precipitating electrons in Jupiter's auroral regions from combined Juno/JADE and Juno/JEDI measurements}
   
   \author{B. Benmahi\inst{1,2} 
          \and
          V. Hue\inst{1}
          \and
          N. Andr\'e\inst{3,4}
          \and
          B. Bonfond\inst{2}
          \and
          M. Blanc\inst{5,3}
          \and
          Z.-Y. Liu\inst{3}
          \and
          B. H. Mauk\inst{6}
          \and
          F. Allegrini\inst{7}
          \and
          G. Clark\inst{6}
          \and
          M. Devinat\inst{3}
          \and
		   G. Gronoff\inst{8,9}
          \and
          T. Le Liboux\inst{3,16}
          \and
          M. Barth\'el\'emy\inst{10,11}
          \and
          C. Lefour\inst{12}
          \and
          T. Cavali\'e\inst{12,13}
          \and
          B. Benne\inst{14,15}
          \and
          T. Gautier\inst{16}
          \and
          T. Briand\inst{1}
          \and
          J. A. Noble\inst{17}
          \and
          J.~A.~Sinclair\inst{18}
}
          
   \institute{Aix-Marseille Université, CNRS, CNES, Institut Origines, LAM, Marseille, France.\\
              \email{bilal.benmahi$@$lam.fr}
        \and
            Laboratory of Planetary and Atmospheric Physics, STAR Institute, University of Liège, Belgium
        \and
            Université de Toulouse, CNRS, CNES, Institut de Recherche en Astrophysique et Planétologie, Toulouse, France 
        \and
            Institut Supérieur de l'Aéronautique et de l'Espace (ISAE SUPAERO), Université de Toulouse, Toulouse, France
        \and
            Weihai Institute for Interdisciplinary Research, Shandong University, Weihai, China
        \and
            Johns Hopkins University Applied Physics Laboratory, Laurel, MD USA
        \and
            Space Science and Engineering Division, Southwest Research Institute, San Antonio, Texas, USA.
        \and
            NASA Langley Research Center, Hampton, Va, USA
        \and
            Science Systems and Applications Inc., Hampton, Va, USA
        \and
            Univ. Grenoble Alpes, CNRS, IPAG, 38000 Grenoble, France
        \and
            Univ. Grenoble Alpes, CSUG, 38000 Grenoble, France
        \and
            Laboratoire d'Astrophysique de Bordeaux, Univ. Bordeaux, CNRS, B18N, all\'ee Geoffroy Saint-Hilaire, 33615 Pessac, France
        \and
            LIRA, Observatoire de Paris, Université PSL, CNRS, Sorbonne Université, Université Paris Cité, 5 place Jules Janssen, 92195 Meudon, France
        \and
            The University of Edinburgh, School of GeoSciences, Edinburgh, UK.
        \and
            Centre for Exoplanet Science, University of Edinburgh, Edinburgh – UK
        \and
            LATMOS/IPSL, CNRS, Sorbonne Université, UVSQ-UPSaclay, Guyancourt, France
        \and   
            Physique des Interactions Ioniques et Moléculaires, CNRS and Aix-Marseille Université, Marseille, France
        \and   
            Jet Propulsion Laboratory, California Institute of Technology, 4800 Oak Grove Drive, Pasadena, CA 91109, USA   
        }

   \date{Received ??/??/??? ; Accepted ??/??/???}

 
\abstract
{Jupiter's auroras are produced by magnetospheric electrons precipitating into its atmosphere. The Juno/JADE and Juno/JEDI instruments together measure these electrons over the energy range $100\,\text{eV}$--$1\,\text{MeV}$. However, their spectra in Jupiter's auroral regions cannot be accurately reproduced by the widely used kappa distribution, hereafter referred to as the classical single-kappa distribution.}
{We aim to statistically characterize the energy flux distributions of precipitating electrons in Jupiter's auroral regions and to derive a new phenomenological energy distribution, referred to as the 4-kappa distribution, better suited to reproduce the combined JADE+JEDI data.}
{We build mean electron energy flux distributions in six auroral sub-regions (the polar, the main, and the outer emission regions, for both hemispheres), combining JADE and JEDI data from PJ3 to PJ35. These data are mapped to the Jovian SIII reference frame using the JRM33 magnetic field model, which accounts for the contribution of the current sheet. We fit these distributions using MCMC in logarithmic space, testing both the classical single-kappa distribution and the 4-kappa distribution.}
{The mean distributions show an overall monotonic power-law decay, with an energy cutoff visible beyond $\sim 500\,\text{keV}$ in most sub-regions, although a fraction of the individual spectra depart from this behaviour at intermediate energies. Mean energies range from $\sim 21$ to $\sim 119\,\text{keV}$, and total energy fluxes range from $\sim 9.5$ to $\sim 28.7\,\text{mW/m}^2$ depending on the sub-region. Fitting with the classical single-kappa distribution fails in all sub-regions, with $\kappa$ parameter systematically converging to the lower bound of the assumed prior range in the MCMC fit. We therefore introduce the 4-kappa distribution, defined as a linear combination of four classical single-kappa distributions modulated by an exponential high-energy cutoff. This distribution accurately fits the spectra in all six auroral sub-regions. At equal total energy flux and mean energy, electron transport simulations show that the 4-kappa distribution deposits its energy at lower altitude than the classical single-kappa distribution, and distributes it differently along the vertical, with direct implications for the UV color ratio, for the atmospheric chemical composition and thermal structure, and for ionospheric conductances.}
{The 4-kappa distribution provides a physically motivated analytical framework for future realistic modeling of electron precipitation and UV auroral emissions at Jupiter.}


   \titlerunning{Precipitating electrons from Juno/JADE and Juno/JEDI data}
   \authorrunning{B. Benmahi et al.}
   \maketitle

\section{Introduction}
\label{introduction}

Jupiter's auroras are the most powerful in the solar system (e.g., \citealt{Broadfoot1979,Clarke1980,Livengood1992,GladstoneSkinner1989,Harris1996}). They result from the coupling between Jupiter's magnetosphere and atmosphere, driven by the precipitation of magnetospheric charged particles into the polar regions (e.g., \citealt{CowleyBunce2001, Hill2001}). Electrons are the dominant precipitating particles. When colliding with atmospheric atomic hydrogen (H) and molecular hydrogen (H$_2$), they produce UV auroral emissions dominated by the H Lyman-alpha line at $121.566\,\text{nm}$ \citep{Broadfoot1979, Dols2000} and the H$_2$ Lyman and Werner bands between $120$ and $170\,\text{nm}$ (e.g., \citealt{Gerard2014, Gustin2016, Benmahi2024a, Benmahi2024b}). In the infrared range, auroral emissions are dominated by the fundamental Q-branch ro-vibrational transitions of H$_3^+$, produced by electron collisions with H$_2$, between $3.9$ and $4\,\mu\text{m}$ \citep{Drossart1989, Trafton1989, Baron1991, Mura2017}. In the visible, these auroras are produced primarily by the collisional excitation of neutral atmospheric species by precipitating electrons, and were first imaged by the SSI camera aboard Galileo \citep{Ingersoll1998, Vasavada1999}.

The magnetosphere-atmosphere coupling also affects the thermal structure of Jupiter's upper atmosphere (e.g., \citealt{Sinclair2017, Sinclair2018, ODonoghue2021}) and its chemical composition (e.g., \citealt{Sinclair2018}), making Jupiter's auroral regions a natural laboratory for studying magnetosphere-atmosphere interactions in giant planets (see review of \citealt{Hue2024}).

To understand and quantify this coupling, the community has combined multi-wavelength observations with physical modeling. UV spectral observations, primarily with HST, have mapped Jupiter's auroral structures with increasing angular and spectral resolution (e.g., \citealt{Clarke1998, Grodent2003, Gerard2013, Gustin2016}). Electron transport models (e.g., \citealt{Grodent2001, Gustin2004, Benmahi2024a, Benmahi2025}) and radiative transfer models (e.g., \citealt{Barthelemy2004, Gustin2013, Benmahi2024a, Benmahi2024b, Benmahi2025}) have been developed to study the relationship between the distributions of precipitating particles and the observed auroral emissions. These precipitated flux distributions directly control the atmospheric energy deposition profile, auroral emission excitation rates, and ionization and dissociation rates. The most direct way to constrain these distributions is in-situ measurement close to the auroral regions.

The first in-situ measurements of energetic particles in Jupiter's magnetosphere were performed during the Pioneer 10 and 11 flybys \citep{VanAllen1974, VanAllen1975}, sensitive to electrons above $\sim 60\,\text{keV}$, followed by Voyager 1 and 2 \citep{Lanzerotti1981}, sensitive to electrons above $\sim 10\,\text{MeV}$, and Galileo \citep{Williams1992}, covering electrons from $15\,\text{keV}$ to $>11\,\text{MeV}$. These missions were, however, limited by their orbital trajectories and did not provide systematic access to the polar auroral regions. NASA's Juno mission \citep{Bolton2017}, which orbits Jupiter on an elliptical polar orbit since July 2016, opened a new era for in-situ measurements of the Jovian magnetospheric plasma, and in particular of precipitating charged particles in the auroral regions. Two instruments are specifically dedicated to the measurement of the energy flux distributions of precipitating particles, such as electrons: JADE (Jovian Auroral Distributions Experiment), covering electron energies from $100\,\text{eV}$ to $100\,\text{keV}$ \citep{McComas2017}, and JEDI (Jupiter Energetic Particle Detector Instrument), sensitive to electrons between $\sim 20\,\text{keV}$ and $1\,\text{MeV}$ \citep{Mauk2017}.

These two instruments have provided an unprecedented characterization of Jupiter's magnetospheric electron distributions. \citet{Mauk2017, Mauk2020} identified the coexistence of monoenergetic distributions, associated with quasi-static electric potential acceleration, and broadband distributions, resulting from stochastic acceleration over a wide energy range. \citet{Clark2018} characterized the energy flux versus mean energy relationships in the main emission loss cone over $30\,\text{keV}$--$1\,\text{MeV}$ using JEDI data only. \citet{Allegrini2020} provided the first comprehensive characterization of Jovian auroral electrons over the full $\sim 100\,\text{eV}$--$1\,\text{MeV}$ range by combining JADE and JEDI data from the first eight perijoves (PJ), and mapping Juno's magnetic footprint onto UV images in both hemispheres. \citet{Paranicas2021} also combined JADE and JEDI data to characterize electron energy spectra near Ganymede's orbit, fitting them with an extended kappa distribution \citep{Hawkins1998}. Their study was however conducted in the context of Jupiter's radiation belt environment at Ganymede's orbital distance, rather than in the Jovian auroral precipitation regions. \citet{Salveter2022} performed a statistical analysis of electron distributions over $30$--$1200\,\text{keV}$ using JEDI data from the first twenty perijoves, and confirmed that broadband distributions dominate in Jupiter's main auroral emission zone.

Monoenergetic, Maxwellian, and kappa distributions (e.g., \citealt{Bonfond2015, Gustin2016, Gerard2016, Benmahi2024a, Benmahi2025}) are commonly used to model the effects of magnetospheric electron precipitation on Jupiter's atmosphere and reproduce observed UV auroral emissions. The kappa distribution \citep{Coumans2002}, with its power-law tail, represents broadband magnetospheric electron populations. However, it was established by fitting electron distributions measured between $\sim 30\,\text{keV}$ and a few hundred keV, the energy range accessible to pre-Juno instruments. Whether it remains valid over the full $100\,\text{eV}$--$1\,\text{MeV}$ range, now accessible through the JADE+JEDI data, is an open question.

Using combined JADE and JEDI measurements from PJ3 to PJ35, we performed a statistical analysis of precipitating electron energy flux distributions in Jupiter's auroral regions over the full $100\,\text{eV}$--$1\,\text{MeV}$ range. Our goals are twofold: first, to statistically characterize the overall shape of these distributions. Second, to establish a new phenomenological distribution, hereafter the 4-kappa distribution, better suited to Juno data, which will serve as a basis for future electron transport and auroral emission modeling. We also compared electron transport in Jupiter's atmosphere using, on one hand, the classical single-kappa distribution (defined in Equation~\ref{eq:kappa_distribution}, Appendix~\ref{model_transport}) and, on the other hand, the 4-kappa distribution introduced in this study.

The paper is organized as follows. We first describe the JADE and JEDI instruments in Section~\ref{observation}, and introduce our method of the data combination in Section~\ref{method}. Results and discussion are presented in Section~\ref{results}, and we conclude in Section~\ref{conclusion}.

\section{JADE and JEDI in-situ measurements}
\label{observation}

\subsection{The JADE and JEDI instruments}

Juno carries two complementary instruments dedicated to in-situ measurements of electron energy flux distributions in Jupiter's auroral regions: JADE (Jovian Auroral Distributions Experiment) and JEDI (Jupiter Energetic Particle Detector Instrument).

JADE consists of three identical electron sensors (JADE-E) and one ion sensor (JADE-I) \citep{McComas2017}. The three JADE-E sensors are mounted $120^{\circ}$ apart around the spacecraft spin axis, providing full pitch angle\footnote{The pitch angle $\alpha$ is the angle between the velocity vector of a charged particle and the local magnetic field direction $\vec{B}$. Precipitating particles have pitch angles smaller than the loss cone angle $\alpha_\mathrm{lc}$, defined by the magnetic mirror condition.} coverage of the electron distributions regardless of the spacecraft rotation phase. JADE-E measures directional differential electron intensities from $100\,\text{eV}$ to $100\,\text{keV}$, with an energy resolution $\Delta E/E$ (FWHM of the energy passband divided by its most probable energy) of $10$--$14\,\%$ and a time resolution of $1\,\text{s}$ \citep{McComas2017, Allegrini2017}. In this study, we use only two of the three JADE-E sensors, as the third was turned off prior to Jupiter orbit insertion and was not operated during the mission.

JEDI is a solid-state detector (SSD) instrument covering electron energies from $\sim 20\,\text{keV}$ to $\sim 1\,\text{MeV}$ \citep{Mauk2017}. It consists of three detector heads (JEDI-90, JEDI-180, and JEDI-270), each equipped with six SSDs providing full pitch angle coverage at a time resolution of $\sim 0.5\,\text{s}$, averaged to $1\,\text{s}$ for statistical reasons \citep{Mauk2017, Clark2018}. We note that the large electron SSD pixel of JEDI-90 failed in April 2023. This does not affect the present study, restricted to PJ3--PJ35 (2016--2021), but should be excluded in analyses of later perijoves.

Together, JADE and JEDI cover a continuous energy range from $\sim 100\,\text{eV}$ to $\sim 1\,\text{MeV}$, with a nominal overlap region between $\sim 20$ and $\sim 100\,\text{keV}$, making their combination a powerful tool for characterizing the full energy flux distributions of precipitating electrons in Jupiter's auroral regions. We note that the upper bound of this overlap depends on JADE's science configuration, which changed over the course of the mission, with the sweep voltage table capping JADE-E at $\sim 50\,\text{keV}$ for part of the mission. It should be noted, however, that the two instruments do not share the same pitch angle coverage at all times. JADE is equipped with electrostatic deflectors that allow it to track the direction of the magnetic field lines, whereas JEDI has a fixed field of view with respect to the spacecraft. This difference in angular sampling can occasionally lead to partial or non-overlapping pitch angle coverage between the two instruments at a given time, a limitation that is taken into account in our spectra merging procedure (Section~\ref{data_merging}).

\subsection{Data and time selection}

We use calibrated differential electron intensity data from JADE and JEDI, covering PJ3 to PJ35, downloaded from the NASA Planetary Data System Planetary Plasma Interactions (PDS PPI) node. The JADE data correspond to level-5 (L5) products, archived under the PDS PPI Data Set ID JNO-J\_SW-JAD-5-CALIBRATED-V1.0. The JEDI data are level-3 (L3) calibrated data, archived under the PDS PPI Data Set ID JNO-J-JED-3-CDR-V1.0 \citep{Mauk2017}. Both datasets provide directional differential intensities, expressed in electrons per $\text{cm}^{-2}\,\text{s}^{-1}\,\text{sr}^{-1}\,\text{keV}^{-1}$, as a function of time and pitch angle. The JADE and JEDI differential intensity distributions are resampled onto a common $1\,\text{s}$ time grid to synchronize both datasets.

For each perijove, we select only the data measured within the time window $(t_\mathrm{ini}, t_\mathrm{fin})$ corresponding to the spectral observations performed by the UVS instrument (UltraViolet Spectrograph; \citealt{Gladstone2017}) aboard Juno. Here $t_\mathrm{ini}$ and $t_\mathrm{fin}$ denote the start and end times of the UVS auroral observation sequence for each perijove. These times are extracted from the UVS data file headers available on the NASA Planetary Data System (Table~\ref{tab:uvs_time_windows}). JADE, JEDI and UVS timestamps are all expressed in ephemeris time (ET) computed from the same SPICE kernels, ensuring that the selected time windows are consistently registered across the three datasets. This selection ensures that the in-situ electron measurements are temporally co-located with the UV auroral observations, enabling future comparisons between the precipitating electron properties derived here and the energy fluxes and mean energies inferred from UVS brightness observations (e.g., \citealt{Benmahi2024a, Vinesse2026}), even though the UVS brightness maps result from the averaging of multiple slit scans over the auroral region while JADE and JEDI provide instantaneous measurements. This time window also restricts the analysis to the portion of each perijove when Juno is close to the planet, thereby limiting contamination from penetrating radiation-belt particles at larger radial distances, and reduces the data volume to a level compatible with our processing pipeline. We note that this temporal window does not introduce a systematic bias in the statistical analysis, as the auroral sub-region coverage remains sufficient across the PJ3 to PJ35 combined data.

\begin{table*}[h!]
\centering
\caption{Time windows $(t_\mathrm{ini}, t_\mathrm{fin})$ of the UVS auroral observation sequences used to select the JADE and JEDI data for each perijove (PJ3--PJ35), as extracted from the UVS data file headers available on the NASA Planetary Data System Atmospheres (ATM) node, Data Set ID JNO-J-UVS-3-RDR-V1.0. Times are given in UTC.}
\label{tab:uvs_time_windows}
\footnotesize
\renewcommand{\arraystretch}{1.2}
\resizebox{\textwidth}{!}{%
\begin{tabular}{lll|lll|lll}
\toprule
PJ & $t_\mathrm{ini}$ (UTC) & $t_\mathrm{fin}$ (UTC) & PJ & $t_\mathrm{ini}$ (UTC) & $t_\mathrm{fin}$ (UTC) & PJ & $t_\mathrm{ini}$ (UTC) & $t_\mathrm{fin}$ (UTC) \\
\midrule
PJ03 & 2016-12-11 12:03 & 2016-12-11 22:02 & PJ14 & 2018-07-16 02:53 & 2018-07-16 10:15 & PJ25 & 2020-02-17 17:35 & 2020-02-17 22:50 \\
PJ04 & 2017-02-02 08:01 & 2017-02-02 17:56 & PJ15 & 2018-09-06 23:43 & 2018-09-07 06:10 & PJ26 & 2020-04-10 12:32 & 2020-04-10 18:45 \\
PJ05 & 2017-03-27 03:52 & 2017-03-27 13:50 & PJ16 & 2018-10-29 20:08 & 2018-10-30 02:04 & PJ27 & 2020-06-02 09:09 & 2020-06-02 15:17 \\
PJ06 & 2017-05-19 01:00 & 2017-05-19 10:59 & PJ17 & 2018-12-21 12:10 & 2018-12-21 21:58 & PJ28 & 2020-07-25 05:30 & 2020-07-25 11:13 \\
PJ07 & 2017-07-10 20:54 & 2017-07-11 06:53 & PJ18 & 2019-02-12 12:40 & 2019-02-12 22:33 & PJ29 & 2020-09-16 01:46 & 2020-09-16 07:08 \\
PJ08 & 2017-09-01 19:48 & 2017-09-02 02:47 & PJ19 & 2019-04-06 11:08 & 2019-04-06 17:12 & PJ30 & 2020-11-08 01:14 & 2020-11-08 06:48 \\
PJ09 & 2017-10-24 12:19 & 2017-10-24 22:41 & PJ20 & 2019-05-29 06:34 & 2019-05-29 13:06 & PJ31 & 2020-12-30 21:30 & 2020-12-31 02:43 \\
PJ10 & 2017-12-16 13:45 & 2017-12-16 22:57 & PJ21 & 2019-07-21 03:04 & 2019-07-21 09:00 & PJ32 & 2021-02-21 16:23 & 2021-02-21 22:38 \\
PJ11 & 2018-02-07 10:14 & 2018-02-07 18:50 & PJ22 & 2019-09-12 02:30 & 2019-09-12 08:38 & PJ33 & 2021-04-15 22:40 & 2021-04-16 04:30 \\
PJ12 & 2018-04-01 06:27 & 2018-04-01 14:44 & PJ23 & 2019-11-03 21:18 & 2019-11-04 03:17 & PJ34 & 2021-06-08 06:56 & 2021-06-08 12:44 \\
PJ13 & 2018-05-24 03:37 & 2018-05-24 10:38 & PJ24 & 2019-12-26 16:40 & 2019-12-26 22:34 & PJ35 & 2021-07-21 07:23 & 2021-07-21 13:13 \\
\bottomrule
\end{tabular}}
\end{table*}

\section{Data processing and combination method}
\label{method}

The goal of this study is to map precipitating electrons in Jupiter's north and south auroral regions using combined JADE and JEDI data, covering the full $100\,\text{eV}$--$1\,\text{MeV}$ energy range. The method proceeds in three steps. First, we project the JADE and JEDI in-situ measurements into the Jovian SIII reference frame, associating each loss cone energy flux distribution with a latitude-longitude position in the auroral regions. Second, we process and combine the spectra from both instruments to produce continuous energy flux distributions over the full energy range. Third, we statistically analyze these observed distributions across three auroral sub-regions, the polar, the main, and the outer emission regions.

\subsection{Mapping precipitating electrons in Jupiter's atmosphere}
\label{data_proj}

Using Juno SPICE kernels, we compute the spacecraft position in the Jovian reference frame as a function of time. We then use the JRM33 magnetic field model \citep{Connerney2020, Connerney2022} to trace the magnetic field line connecting the spacecraft to Jupiter's atmosphere at each time step, giving Juno's magnetic footprint in the SIII reference frame at the 1 bar pressure level altitude. Juno follows a polar orbit around Jupiter, and during each PJ passes a few thousand kilometers above the north and south auroral regions, so its magnetic footprint crosses the polar regions including the auroral zones.

The JADE and JEDI in-situ measurements are associated with pitch angle as a function of time. At each time step, we compute the magnetic field magnitude $|\vec{B}|$ at the spacecraft altitude and at the $1\,\text{bar}$ atmospheric pressure level to derive the loss cone angle, and we select only the electron distributions whose pitch angle falls within the downward loss cone\footnote{This selection is based on the pitch angle of the centroid direction of each detector's field of view, not on its full angular extent. Given JEDI's $9^\circ \times 17^\circ$ (full width at half maximum, FWHM) field of view, part of a given viewing direction may lie outside the loss cone even when its centroid lies inside, which can occasionally affect individual measurements.}. Each precipitating electron distribution is then associated with a latitude-longitude coordinate corresponding to Juno's magnetic footprint in Jupiter's atmosphere.

\begin{figure*}[h!]
    \centering
    \includegraphics[width=9cm, keepaspectratio]{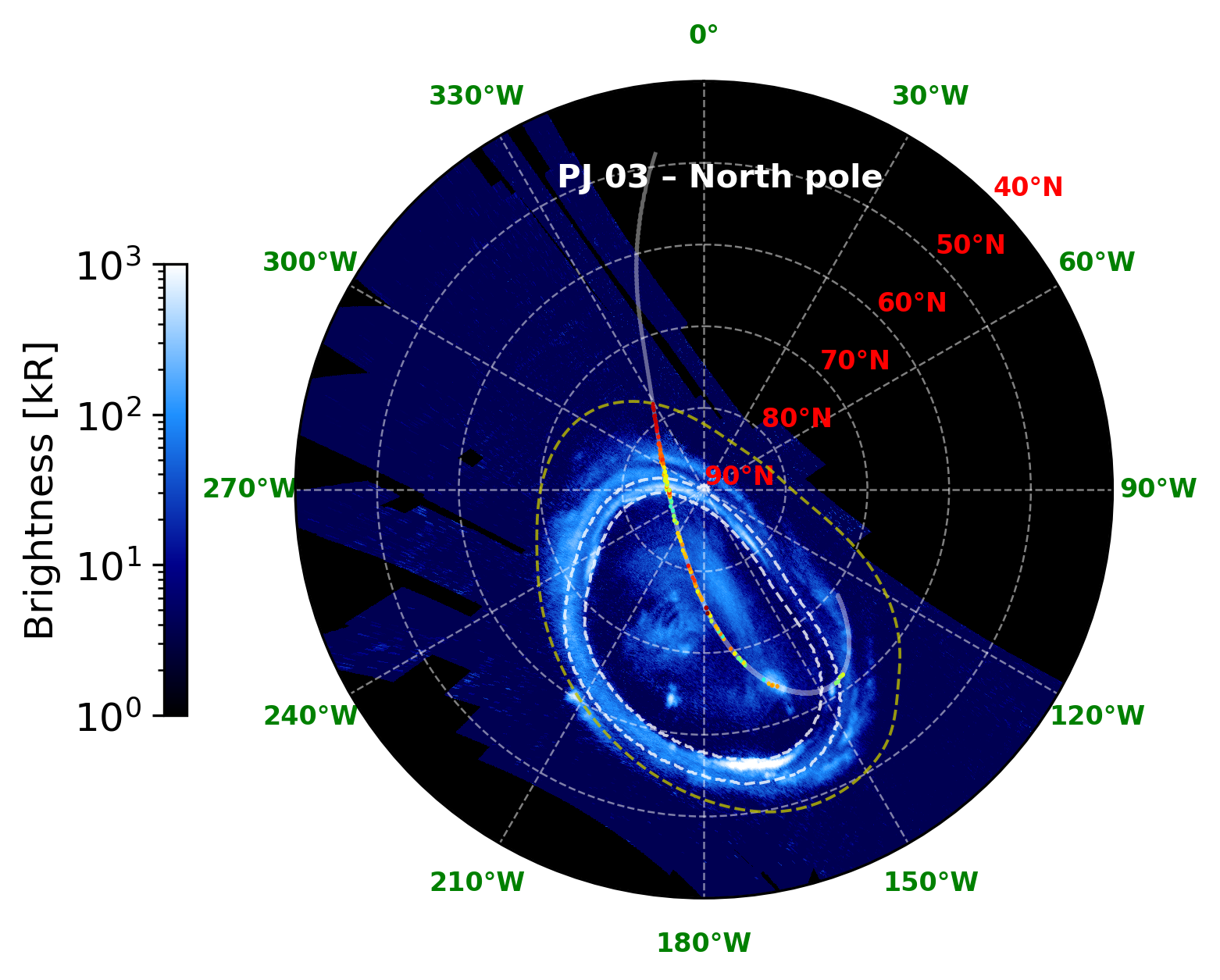}
    \includegraphics[width=9cm, keepaspectratio]{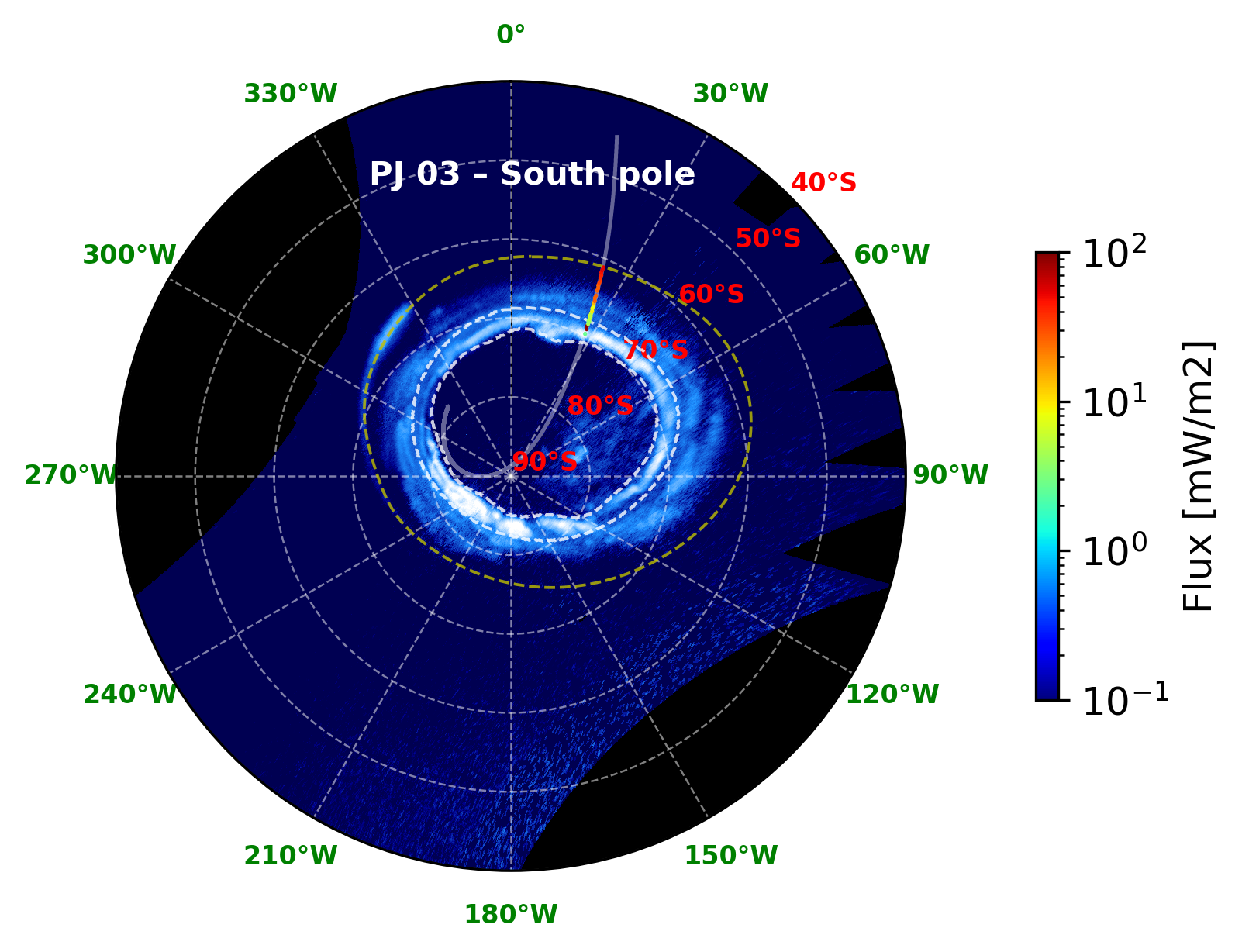}
    \caption{Projection of Juno's magnetic footprint over the north (left) and south (right) Jovian auroral regions during PJ3. The color along the track (in gray) indicates the precipitating energy flux measured by JADE and JEDI at each latitude-longitude position in SIII coordinates.}
    \label{fig:PJ3_Juno_footprint}
\end{figure*}

The JADE and JEDI time resolution is $1\,\text{s}$, corresponding to a displacement of Juno's magnetic footprint of about $40\,\text{km}$ at the Jovian atmospheric level (at the cloud level, $1\,\text{bar}$). We sample the SIII coordinates on a $0.33^\circ \times 0.33^\circ$ latitude-longitude grid, corresponding to a spatial sampling of about $400\,\text{km} \times 400\,\text{km}$. In each grid cell crossed by Juno's magnetic footprint, we compute a mean precipitating electron energy flux distribution by averaging all individual measurements falling within that cell. As Juno moves along its trajectory, successive measurements within a given cell probe slightly different magnetic field lines, so this averaging reflects both the reduction of instrumental counting noise and the natural spatial and temporal variability of the precipitating electron population within the cell. The resulting mean distribution therefore provides a statistically representative estimate of the local precipitating electron energy flux. Figure~\ref{fig:PJ3_Juno_footprint} shows an example of Juno's magnetic footprint projection during PJ3 over the north and south Jovian auroral regions, with the precipitating energy flux overlaid along the footprint track at each latitude-longitude position.

The energy flux $\mathcal{F}$ (in $\text{mW}\,\text{m}^{-2}$) is computed from the JADE+JEDI precipitating energy flux distribution $f(E)$ by integrating over the full energy range:
\begin{equation}
\mathcal{F} = \int E \cdot f(E)\, \mathrm{d}E,
\label{eq:energy_flux}
\end{equation}
where $E$ is the electron energy and $f(E)$ is the precipitating energy flux distribution. Assuming that the directional differential intensity $h(E)$, expressed in $\text{electrons}\,\text{cm}^{-2}\,\text{s}^{-1}\,\text{keV}^{-1}\,\text{sr}^{-1}$, is isotropic within the loss cone and conserved along the field line, and accounting for the projection effect and for the opening of the flux tube as particles precipitate toward the atmosphere, $f(E)$ is given by $f(E) = \pi\, h(E)$, independently of the loss cone half-angle.

\subsection{Definition of the auroral sub-regions}
\label{auroral_regions}

We distinguish three auroral sub-regions based on boundaries established in the literature (e.g., \citealt{Groulard2024, Vinesse2026}). The main emission region (ME) is bounded by the white dashed contours visible in Figure~\ref{fig:PJ3_Juno_footprint}, derived by \citet{Groulard2024} from Juno/UVS observations for each PJ. The polar emission region (PE) is defined as the area inside the inner boundary of the ME. The outer emission region (OE) covers the area between the outer boundary of the ME and the mean magnetic footprint of Io, represented by the yellow dashed contours in the same figure, and established using the JRM33 magnetic field model of Jupiter accounting for the contribution of the magnetic field generated by the current sheet \citep{Connerney2020, Connerney2022}. We emphasize that, despite its name, the OE region maps to lower latitudes than the PE region and is therefore associated with the middle magnetosphere, whereas the PE region, located poleward of the ME, maps to the more distant magnetosphere.

\subsection{Correction of the Minimum Ionizing Bump artifact in JEDI data}
\label{data_mib}

JEDI data contain an instrumental artifact known as the Minimum Ionizing Bump (MIB), which appears as a spurious peak centered at $\sim 160\,\text{keV}$ in the electron spectra. This artifact is caused by energetic electrons penetrating the solid-state detectors themselves \citep{Mauk2017, Mauk2026}. Following the same general approach as \citet{mauk_diverse_2018} for correcting the MIB peak, we fit the spectral continuum between $\sim 100\,\text{keV}$ and $\sim 280\,\text{keV}$ with a polynomial smoothing function, extrapolate the underlying differential intensity beneath the peak, and subtract it. The method differs from \citet{mauk_diverse_2018} only in the specific functional form used for the continuum fit. This correction is restricted to the MIB feature itself, and the high-energy tail beyond $\sim 280\,\text{keV}$ is left unmodified. As with the approach of \citet{mauk_diverse_2018}, this may also remove small spectral structures of physical origin, such as monoenergetic contributions to the global distribution. This is not a limitation here, since this study focuses exclusively on the continuum of the global spectral distribution.

\subsection{JADE and JEDI spectra processing and combination}
\label{data_combinaison}

JEDI spectra from PJ20 onward show a systematic anomalous increase in differential intensity at low energies, below $\sim 50\,\text{keV}$. This anomaly originates from a known processing error affecting the ''small pixel'' data at the lowest energies, which has since been corrected in the most recent JEDI calibration matrices but remains present in the archived PDS data used in this study (B. Mauk, private communication). At the time of writing, the corrected calibration matrices have not yet been publicly released or archived on the PDS, and the JEDI team is still in the process of reprocessing the full dataset. We could therefore not apply this correction directly and instead reconstructed the affected low-energy portion of each spectrum as described below. Precipitating electron spectra at Jupiter are well described by a power-law continuum over this energy range (e.g., \citealt{Clark2018, Salveter2022}), so we replace the affected low-energy portion of each spectrum with a power-law extrapolation, whose exponent is determined from a log-log linear fit to the unaffected part of the spectrum in the $[50\,\text{keV}$--$160\,\text{keV}]$ range.

Finally, to exclude spectra that are too noisy, we apply a spectral quality criterion based on the JEDI spectra only, since JADE spectra are less noisy. To compute this criterion, we first fit the continuum of each spectrum with a second-order polynomial in log-log space above $E_\mathrm{min} = 40\,\text{keV}$. This quality criterion, hereafter denoted $Q$, is defined as the ratio of the mean absolute value of the fitted continuum to the standard deviation of the residuals between the observed spectrum and this continuum, both computed in logarithmic space. This ratio does not correspond to a true instrumental signal-to-noise ratio, but rather quantifies how closely the observed spectrum follows the smooth power-law continuum previously reported for Jovian auroral electron spectra \citep{Allegrini2020}. This criterion is based solely on the shape of the calibrated intensity spectrum and does not use the raw counts-per-accumulation values provided in the JEDI Level 3 data. The threshold $Q \geq 5$ was chosen empirically, after a systematic comparison of spectra above and below this value showed that it reliably separates visibly noisy, unusable spectra from those of acceptable quality.

\subsection{JADE and JEDI spectra merging}
\label{data_merging}

Before merging, the precipitating energy flux distribution of each instrument is computed at each time step $t$ by averaging over all viewing directions whose pitch angle falls within the downward loss cone. For JEDI, this selection applies to the $N_\mathrm{JEDI}$ valid viewing directions among the $3 \times 6 = 18$ available directions (3 detector heads $\times$ 6 directions), satisfying $\alpha > 180^{\circ} - \alpha_\mathrm{lc}(t)$ in the northern hemisphere, or $\alpha < \alpha_\mathrm{lc}(t)$ in the southern hemisphere, where $\alpha_\mathrm{lc}(t)$ is the loss cone angle at time $t$, derived from the magnetic field magnitude at the spacecraft and at the $1\,\text{bar}$ pressure level as described in Section~\ref{data_proj}. For JADE, the same selection applies to the $N_\mathrm{JADE}$ valid directions among the $2 \times 16 = 32$ available directions (2 sensors $\times$ 16 viewing directions). The mean precipitating differential flux of each instrument at time $t$ is then:
\begin{equation}
\bar{f}_\mathrm{inst}(E, t) = \frac{\pi}{N_\mathrm{inst}}\sum_{k=1}^{N_\mathrm{inst}} h_k(E, t),
\end{equation}
where $h_k(E, t)$ is the directional differential intensity measured in direction $k$.

The resulting $\bar{f}_\mathrm{JADE}$ and $\bar{f}_\mathrm{JEDI}$ spectra are merged over a narrower sub-range of the nominal overlap energy domain, between $\sim 20\,\text{keV}$ and $\sim 30\,\text{keV}$, since JADE measurements become occasionally unreliable above $\sim 30\,\text{keV}$, close to the upper edge of its energy range. This merging is performed by spectrally averaging the energy flux distributions of both instruments to produce a continuous transition between the two energy ranges.

Some spectra show a systematic flux offset between JADE and JEDI in the overlap energy range, with an absolute mean ratio between the two instruments ranging from $2$ to $10$. This significantly exceeds the typical agreement of ${\sim}0.8$--$1.2$ observed under optimal measurement conditions (\citealt{Allegrini2020}, Text S2 and Figure S5), and cannot be attributed to calibration uncertainties alone. It most likely results from the different fields of view of JADE and JEDI, which do not always sample the same directions at a given time, and can lead to partial sampling of the precipitating electron distribution, particularly in the presence of narrow field-aligned beams (\citealt{Allegrini2020}, Figure S6). To identify and exclude these cases, we adopt a selection criterion based on the level of agreement between the two instruments. At a given time $t$, denoting $\bar{f}_\mathrm{JADE}$ and $\bar{f}_\mathrm{JEDI}$ as the mean intensities of the two instruments in the overlap energy range, we compute the ratio $R(t) = \bar{f}_\mathrm{JEDI}(t) / \bar{f}_\mathrm{JADE}(t)$ and retain only the spectra for which $0.8 \leq R(t) \leq 1.2$, matching the agreement range identified by \citet{Allegrini2020} under optimal measurement conditions. Spectra falling outside this range, for which the disagreement between the two instruments cannot be reliably resolved, are not considered in our analysis.

\section{Results and discussion}
\label{results}

We present the results of the statistical analysis of precipitating electron energy flux distributions in Jupiter's auroral regions, obtained by combining JADE and JEDI data from PJ3 to PJ35. We first illustrate the projection and spectra combination method on a single case, PJ3, before extending the analysis to all perijoves. We then present the mean energy flux distributions obtained for each of the three auroral sub-regions, along with the method used to fit them by phenomenological distributions and the resulting best-fit parameters (Section~\ref{fit_of_the_mean_distributions}). Finally, we compare electron transport in Jupiter's atmosphere using the classical single-kappa distribution on one hand, and the 4-kappa distribution introduced in this study on the other hand.

\subsection{JADE and JEDI data combination for PJ3}
\label{JADE_JEDI_PJ3}

\begin{figure*}[h!]
    \centering
    \includegraphics[width=9cm, keepaspectratio]{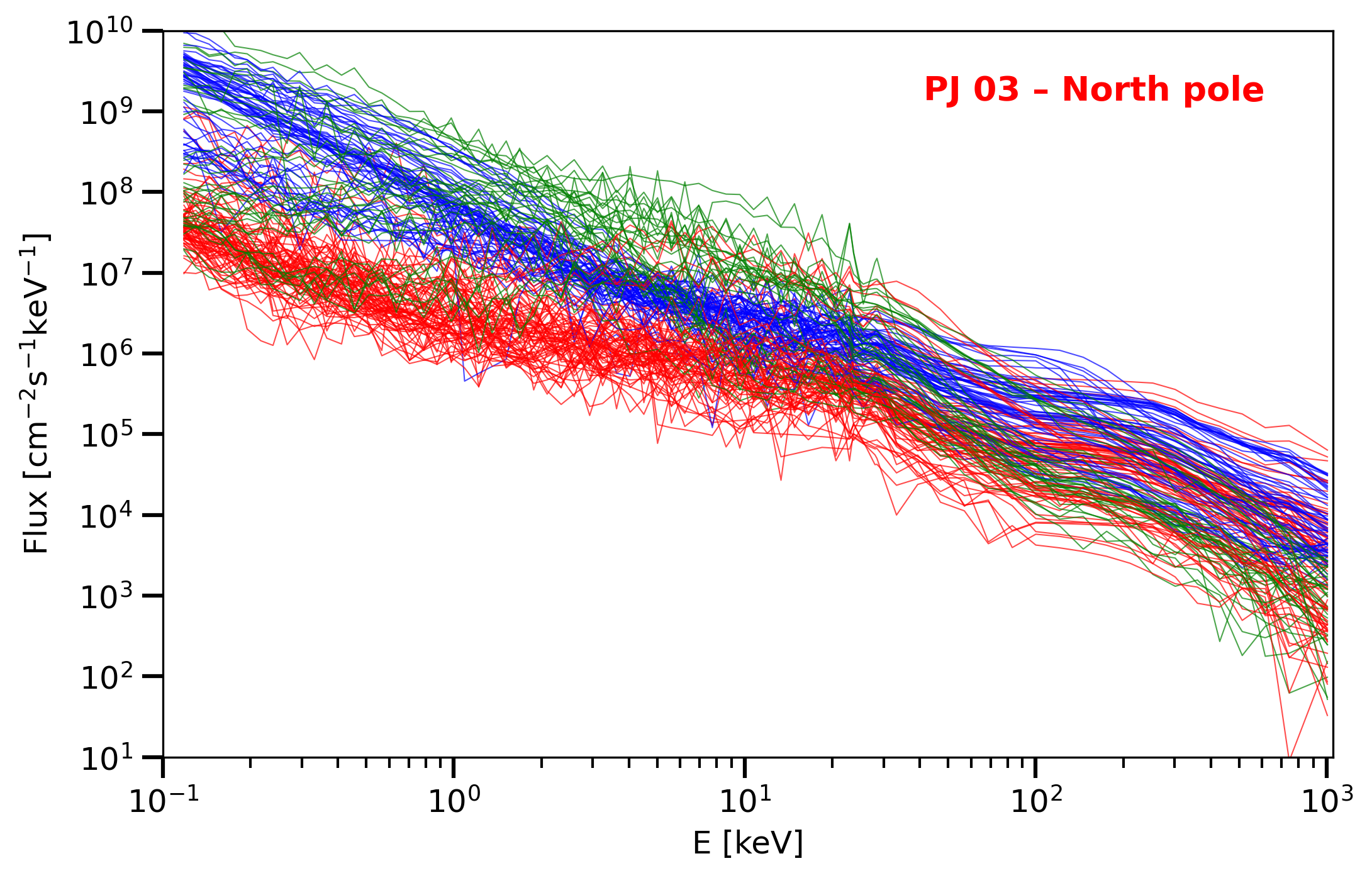}
    \includegraphics[width=9cm, keepaspectratio]{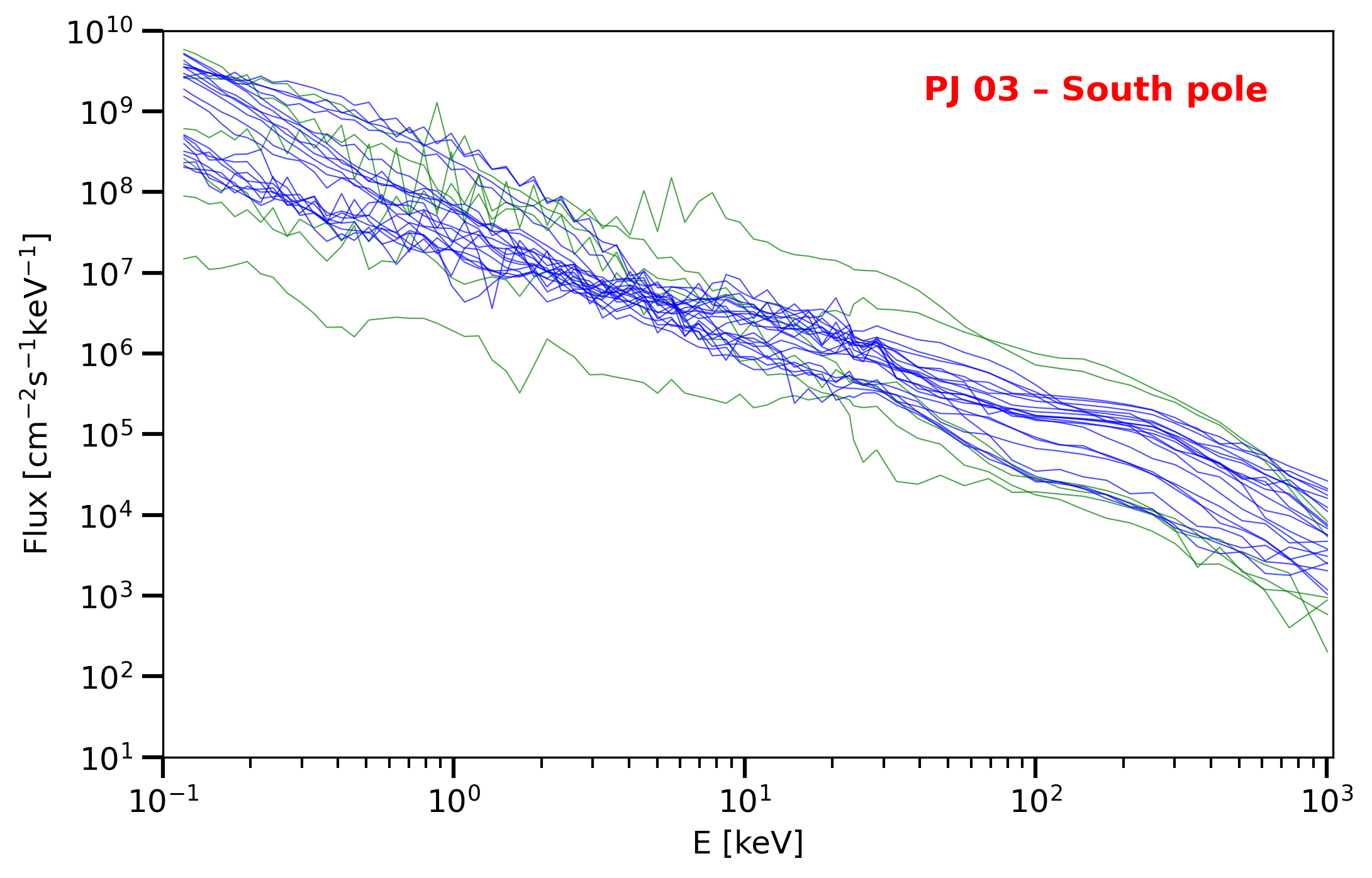}
    \caption{Precipitating electron energy flux distributions measured along Juno's magnetic footprint in the north (left) and south (right) Jovian auroral regions during PJ3. Red, green, and blue curves correspond to the spectra measured in the PE, ME and OE, respectively, as defined in Section~\ref{auroral_regions}.}
    \label{fig:spectra_PJ3_Juno}
\end{figure*}

Figure~\ref{fig:PJ3_Juno_footprint} shows Juno's magnetic footprint track during PJ3 over both Jovian polar regions, overlaid on UV brightness maps from Juno/UVS for the same PJ. These brightness maps were published by \citet{Vinesse2026} for the first 69 perijoves\footnote{The brightness map database is available at \url{https://doi.org/10.58119/ULG/SRUBKL}}. In both hemispheres, the Juno footprint successively crosses the three auroral sub-regions defined in Section~\ref{auroral_regions}. A brightness peak visible at $\sim 65^{\circ}$N and $\sim 160^{\circ}$W in the PE region appears to spatially coincide with an energy flux peak along the footprint track. 
This coincidence should be interpreted with caution, however, since the UVS brightness maps result from averaging multiple slit scans over the auroral region, while JADE and JEDI provide instantaneous measurements. 
In the northern hemisphere, no notable energy flux increase is observed when the footprint crosses the ME region, whereas in the south a flux peak appears to coincide with the ME region. This illustrates the complexity of the coupling between in-situ precipitating particle measurements and time-integrated auroral UV observations.

Figure~\ref{fig:spectra_PJ3_Juno} shows the precipitating electron energy flux distributions measured along Juno's magnetic footprint during PJ3. Red, green, and blue spectra correspond to the PE, ME, and OE regions, respectively. Energy flux intensity differences between the auroral sub-regions do not show a simple systematic dependence on spacecraft altitude during the perijove. Rather, the observed spectral amplitudes appear more randomly distributed among the PE, ME, and OE regions. Overall, the precipitating electron distributions in all auroral regions show similar shapes, characterized by a monotonic power-law decay and an energy cutoff at the upper limit of the JEDI energy window. We note that some of the amplitude variability could in principle reflect ongoing particle acceleration along the field line at the time of measurement. However, since this study focuses on the global shape of the precipitating electron energy flux distributions rather than on individual, possibly transient events, such acceleration episodes do not affect the statistical characterization presented here.

These electron spectral shapes have already been observed by \citet{Allegrini2020} in their characterization of the energy flux and mean energy of precipitating electrons in the ME region from the first eight perijoves. In this auroral sub-region, the characteristics of the precipitating electron distributions obtained with the combined JADE and JEDI data in our study, including the low- and high-energy amplitudes, spectral shape, energy flux, and mean energy, are in good visual agreement with the mean spectra shown in Figure 4 of \citet{Allegrini2020}.

Finally, more distributions are measured in the northern hemisphere than in the southern hemisphere during PJ3, particularly in the PE region. This imbalance results primarily from the requirement that both JADE and JEDI provide a valid, well-overlapping spectrum at a given time (Section~\ref{data_merging}). During this specific perijove, this simultaneous coverage condition happens to be met less often in the southern PE region than in the north. This PJ-to-PJ imbalance is not systematic, as the northern and southern hemispheres are more evenly sampled during other perijoves.

\subsection{JADE and JEDI data combination from PJ3 to PJ35}
\label{JADE_JEDI_all_PJs}

\begin{figure*}[h!]
    \centering
    \includegraphics[width=9cm, keepaspectratio]{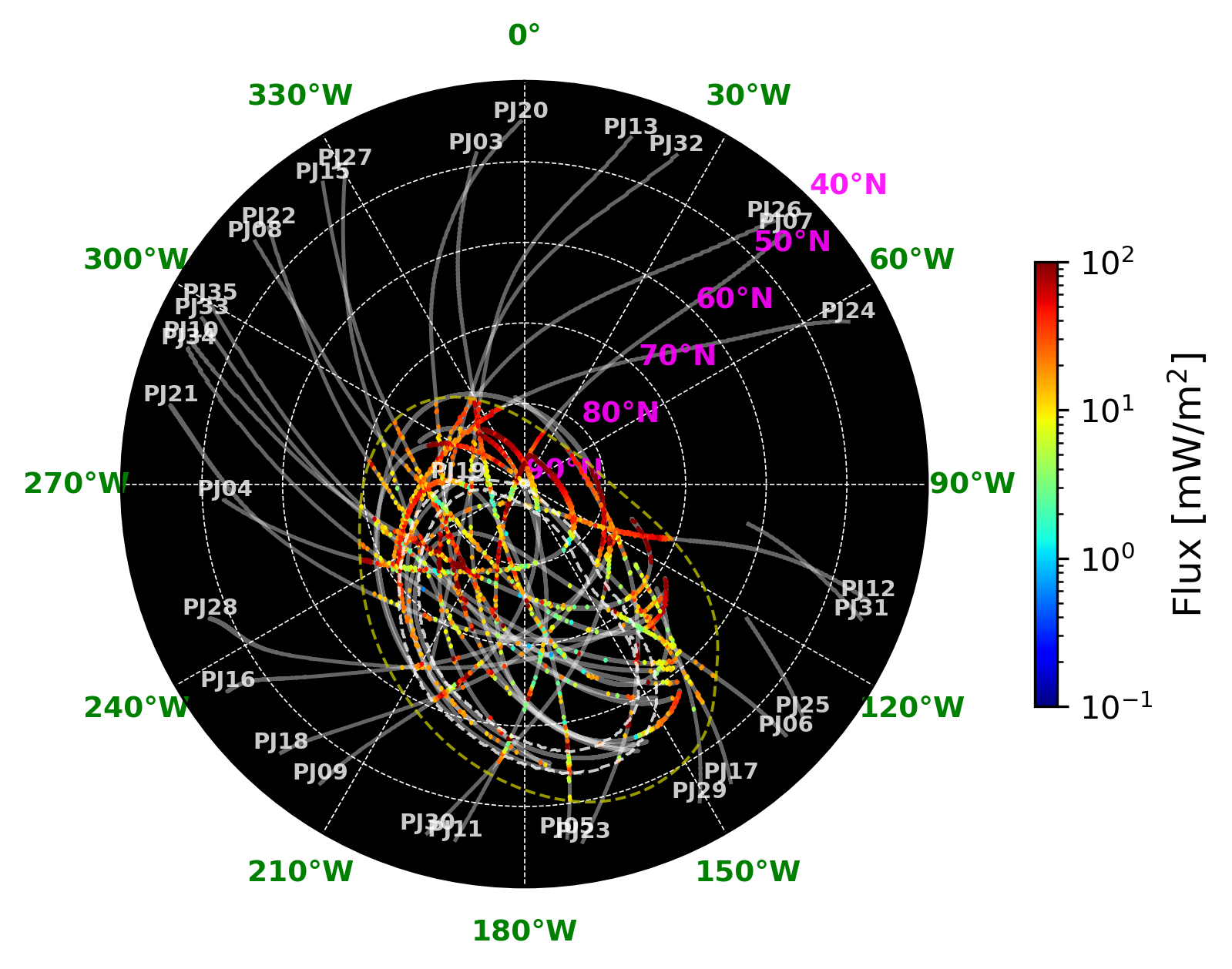}
    \includegraphics[width=9cm, keepaspectratio]{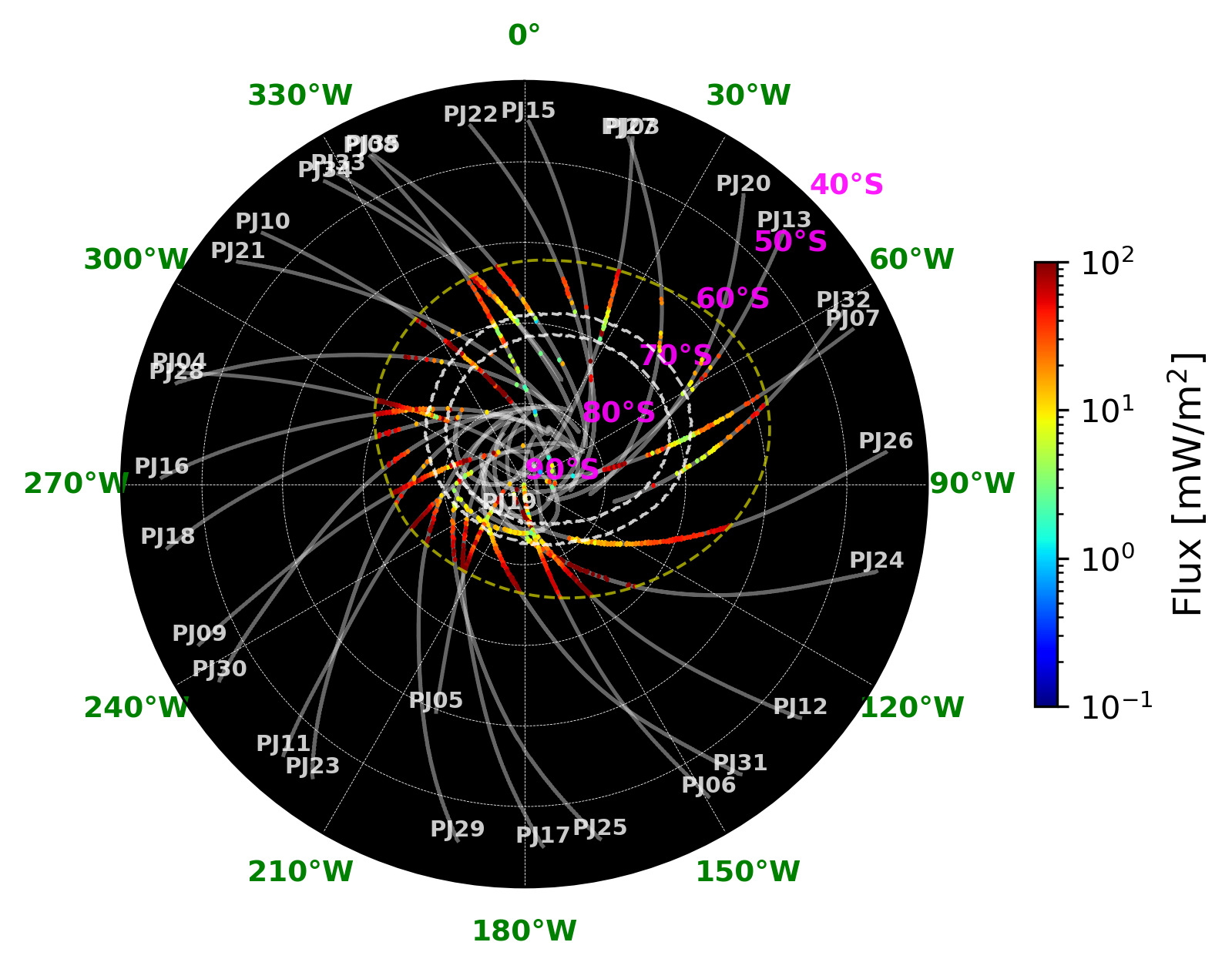}
    \caption{Projection of Juno's magnetic footprint over the north (left) and south (right) Jovian auroral regions combining data from PJ3 to PJ35. The colors along the Juno footprints indicate the precipitating energy flux measured by JADE and JEDI at each latitude-longitude position in SIII coordinates.}
    \label{fig:all_PJ_Juno_footprint}
\end{figure*}

The goal of this study is to build mean precipitating electron energy flux distributions in each of the three auroral sub-regions defined in Section~\ref{auroral_regions}, for both the northern and southern hemispheres. These mean distributions are not intended to represent any individual precipitation event, but rather to statistically characterize the global spectral shape of precipitating electrons in each auroral sub-region, in order to derive a phenomenological distribution suited to future electron transport modeling (Appendix~\ref{model_transport}). As shown in Figures~\ref{fig:PJ3_Juno_footprint} and \ref{fig:spectra_PJ3_Juno}, JADE+JEDI spectra from a single perijove adequately sample the PE and OE regions in terms of spatial coverage. However, Juno's magnetic footprint crosses the ME region very briefly during each perijove, providing too few spectra to build a statistically robust mean distribution there. To address this, we combine JADE+JEDI data from PJ3 to PJ35.

Figure~\ref{fig:all_PJ_Juno_footprint} shows the superposition of Juno's magnetic footprint tracks for PJ3 to PJ35 in both hemispheres. The spatial coverage of all auroral sub-regions is significantly improved compared to a single perijove. Only the southern PE region shows more limited coverage, consistent with the same JADE-JEDI simultaneous overlap requirement discussed in Section~\ref{JADE_JEDI_PJ3}, which is not met as often in this specific sub-region across the combined set of perijoves. This does not affect our analysis, as the number of available spectra remains sufficient to build robust mean distributions in all sub-regions.

We note that, in Figure~\ref{fig:all_PJ_Juno_footprint}, the precipitating energy flux appears to increase markedly as the footprint track approaches the outer boundary of the OE region, close to the mean magnetic footprint of Io. This enhancement is not associated with a corresponding brightening in the Juno/UVS images, and is therefore unlikely to reflect a genuine increase in precipitating electron flux at these low latitudes. It more plausibly results from contamination by penetrating energetic electrons from Jupiter's radiation belts, which JADE and JEDI are not designed to fully exclude at these locations. No spatial filter is applied to exclude spectra close to this outer boundary, so the energy flux values shown at the lowest latitudes, near the OE outer boundary, should be interpreted with caution. This contamination is expected to affect primarily the total energy flux $\mathcal{F}_\mathrm{tot}$ of the OE region rather than the shape of the mean distribution, since this study focuses on the spectral shape of precipitating electrons rather than on their absolute amplitude.

\citet{Salveter2022} performed a similar analysis by mapping precipitating electrons in the ME region using only JEDI data from the first 20 PJs. Their results show that less than 7\% of precipitating electron distributions have monoenergetic structures, while more than 90\% follow broadband distributions, resulting from stochastic acceleration in the magnetosphere prior to precipitation. Our study focuses on characterizing this broadband continuum using the combined JADE and JEDI data, covering a broader energy range and extending the analysis to all six auroral sub-regions. As described in Section~\ref{data_mib}, our MIB correction procedure can also remove localized monoenergetic structures superimposed on the continuum in the $\sim 100$--$280\,\text{keV}$ range. The mean distributions presented below therefore characterize the broadband continuum of precipitating electrons rather than any individual monoenergetic component, consistent with the dominant broadband population identified by \citet{Salveter2022}. The main differences between sub-regions lie in their mean energies and amplitudes.

\begin{figure*}[h!]
    \centering
    \includegraphics[width=9cm]{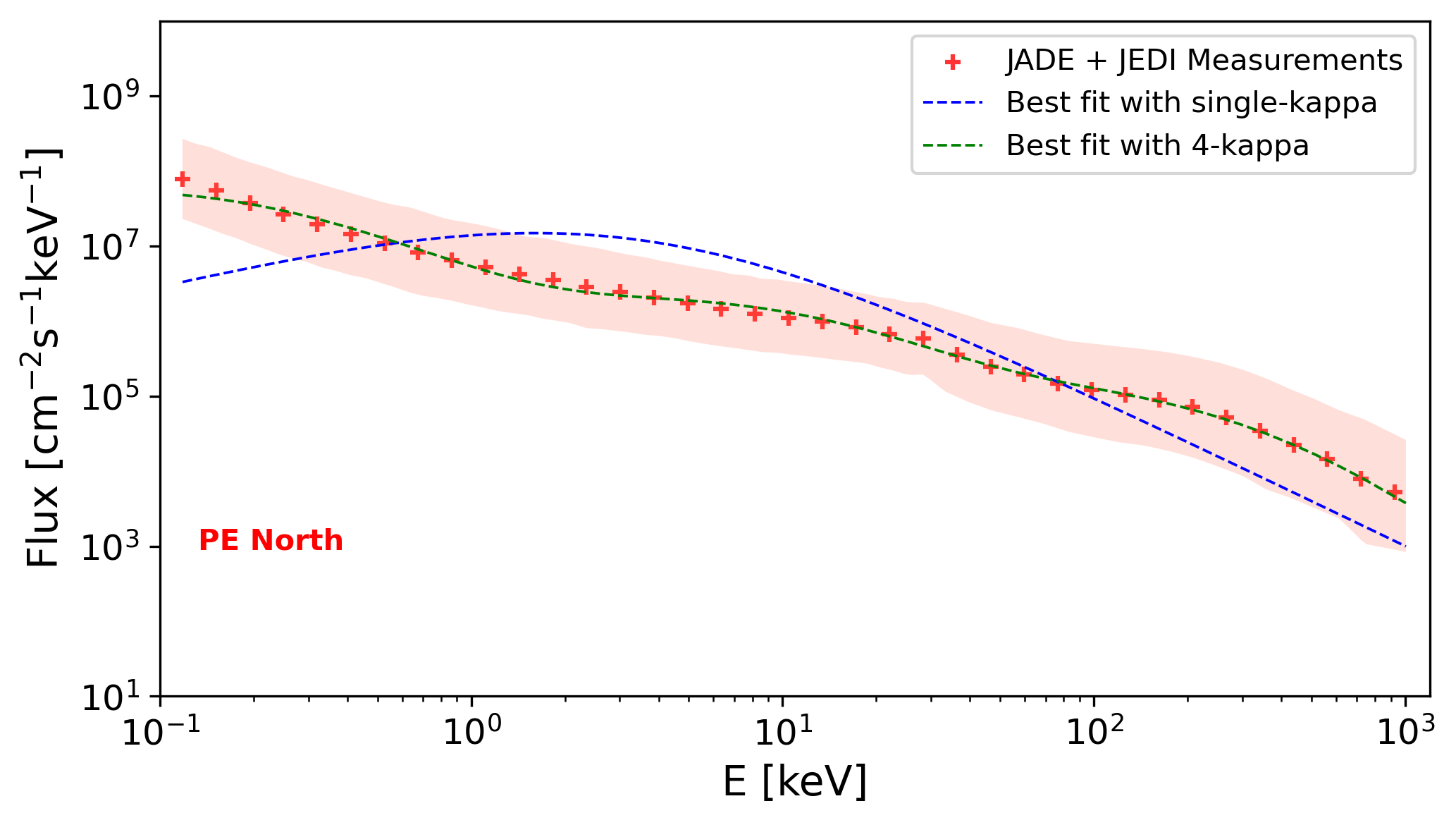}
    \includegraphics[width=9cm]{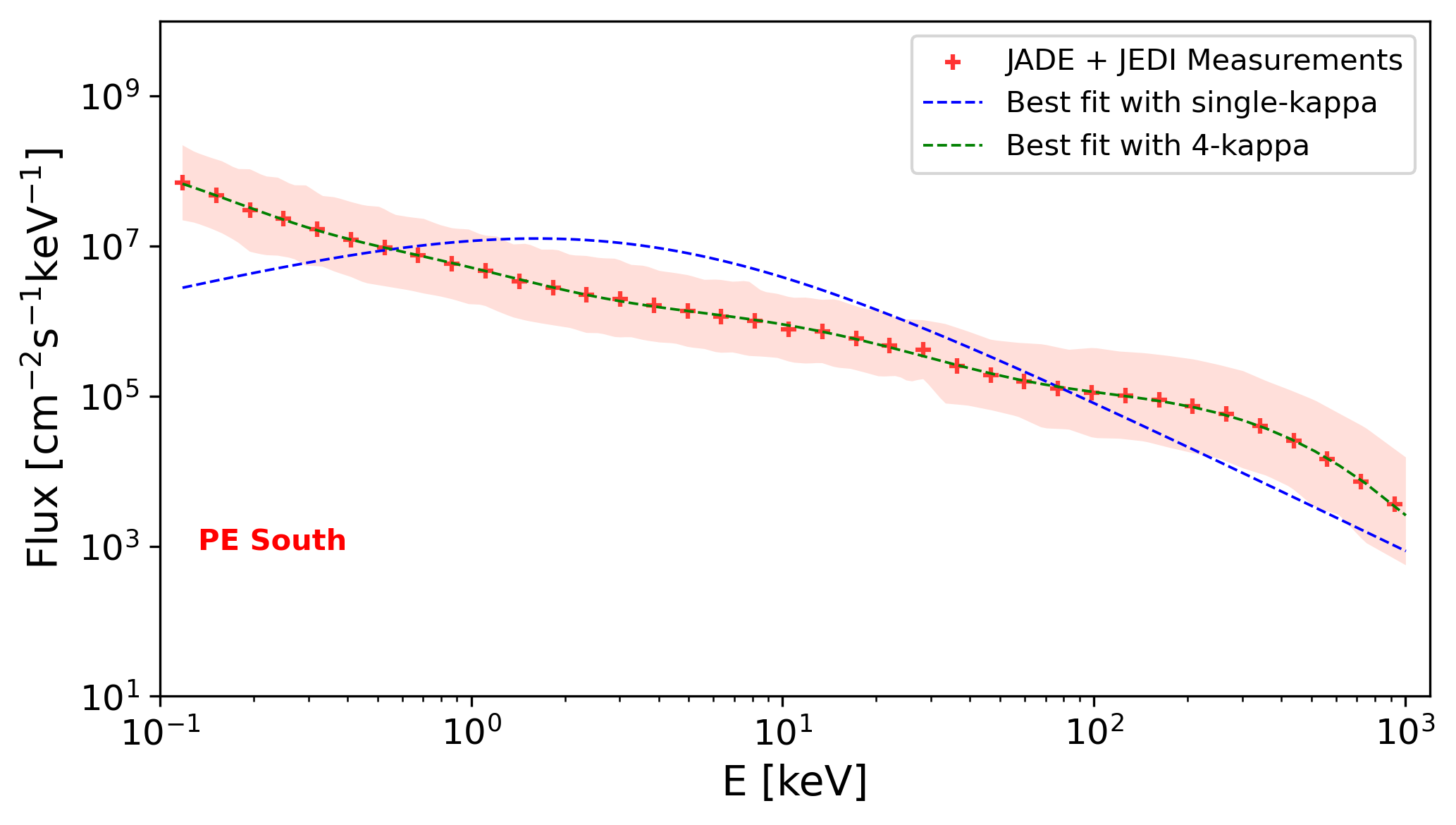}
    \includegraphics[width=9cm]{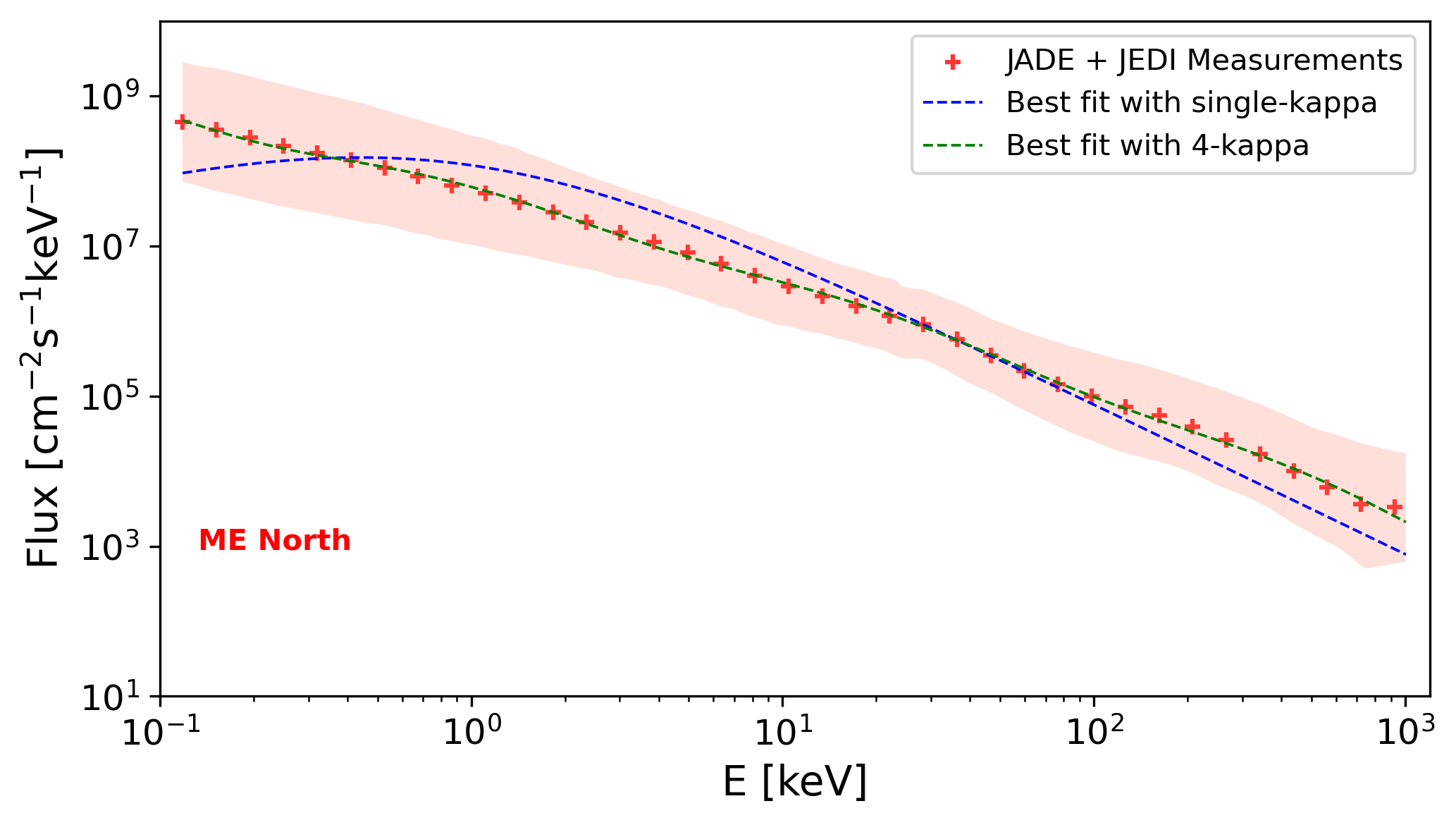}
    \includegraphics[width=9cm]{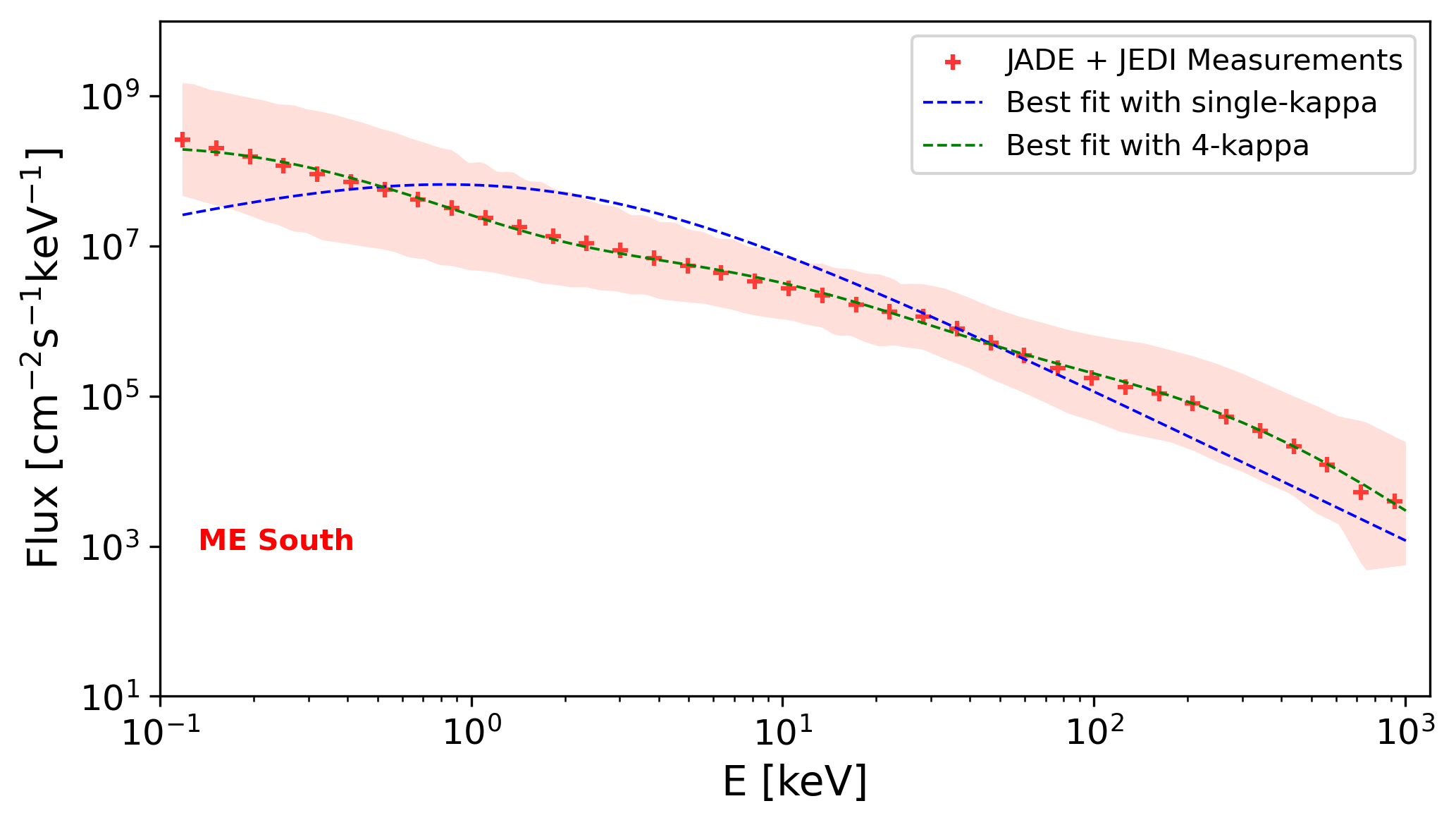}
    \includegraphics[width=9cm]{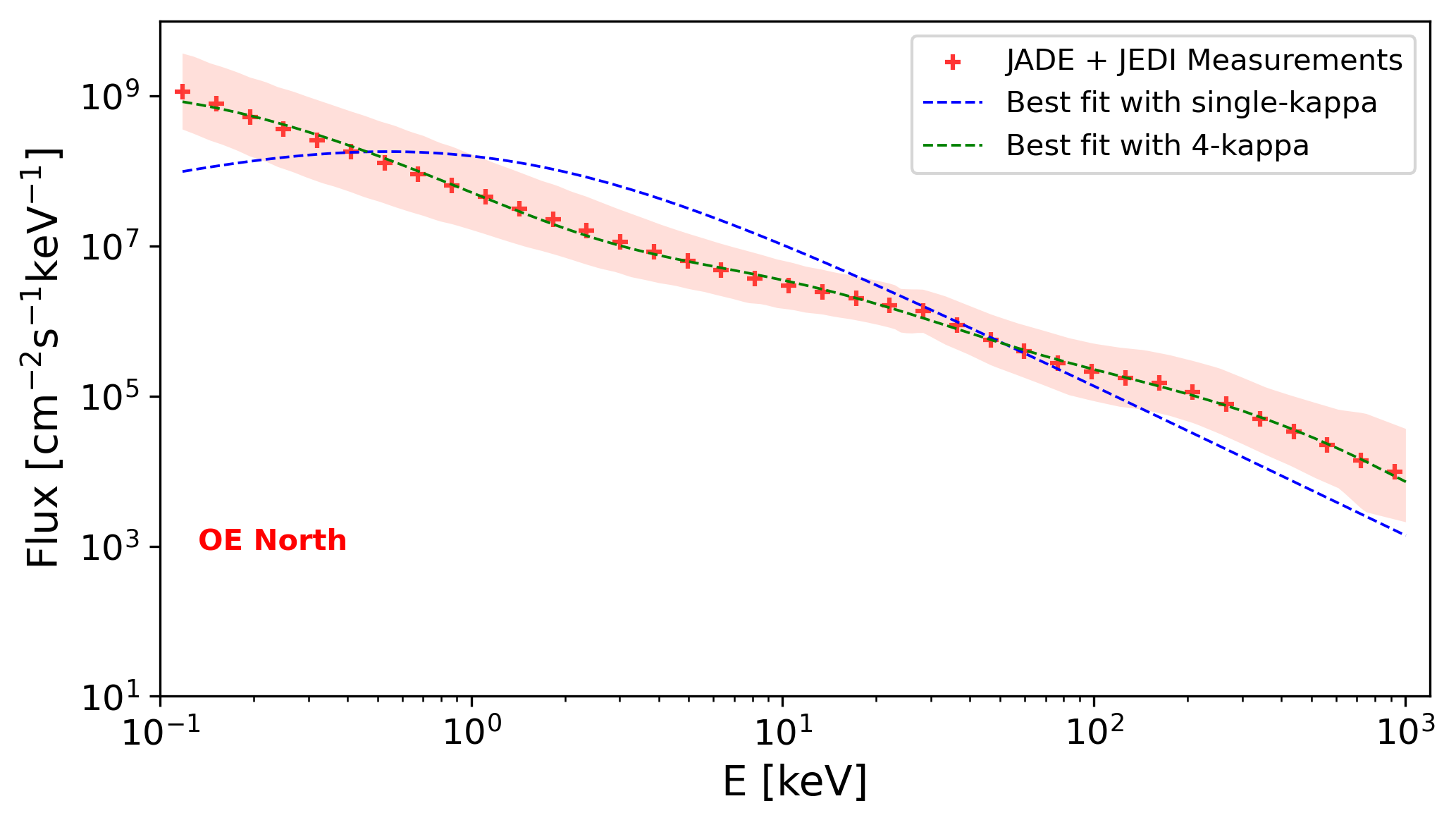}
    \includegraphics[width=9cm]{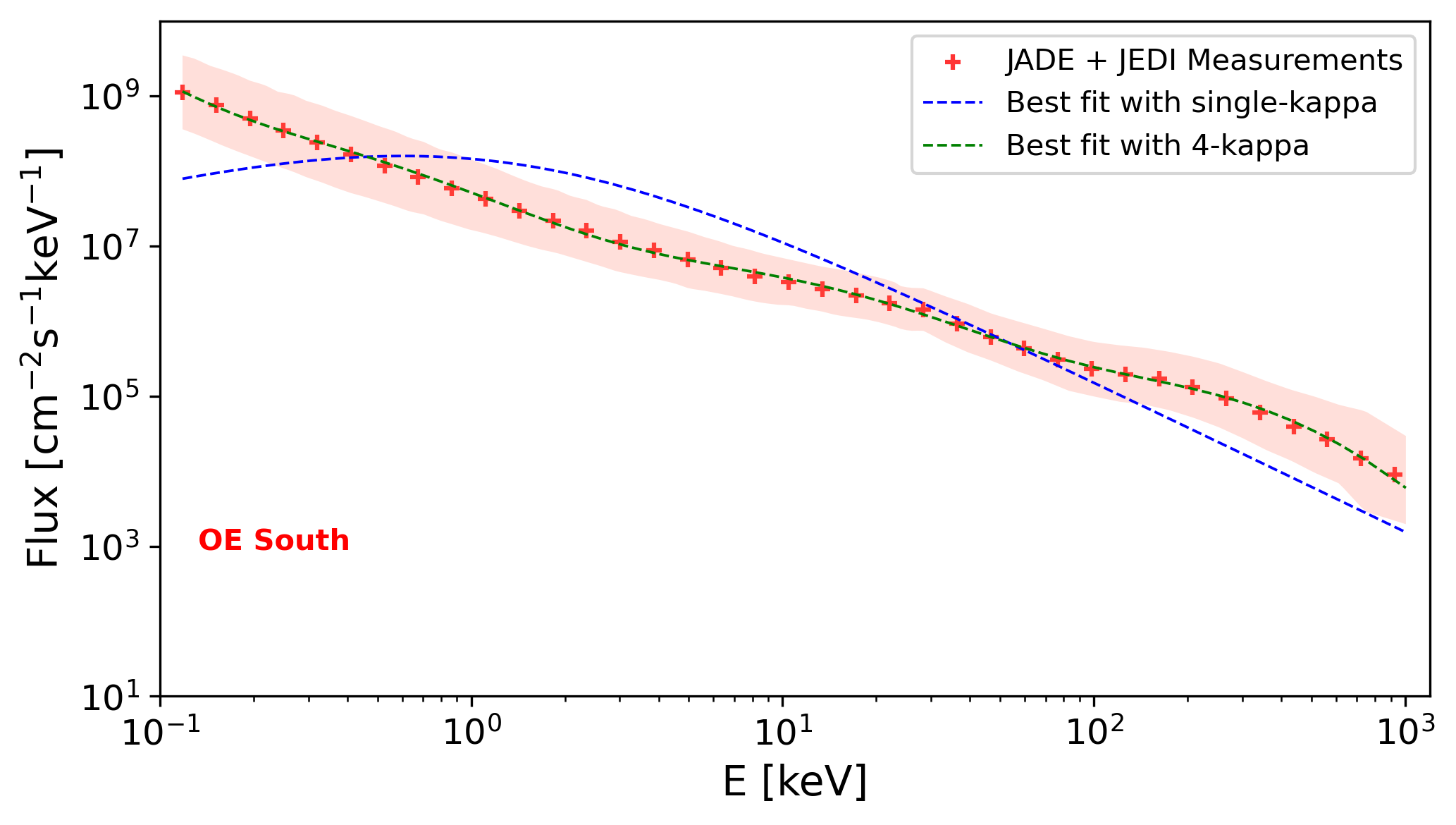}
    \caption{Mean precipitating electron energy flux distributions in the PE, ME, and OE regions, as defined in Section~\ref{auroral_regions}. From top to bottom, the panels show the northern and southern PE regions, the northern and southern ME regions, and the northern and southern OE regions. The shaded band represents the $1\sigma$ confidence interval computed in logarithmic space around each mean spectrum.}
    \label{fig:mean_spectra_JADE_JEDI}
\end{figure*}

Figure~\ref{fig:mean_spectra_JADE_JEDI} shows the mean precipitating electron energy flux distributions obtained in the six auroral sub-regions, together with the $1\sigma$ confidence band computed in logarithmic space to quantify the spread of individual measurements around the mean. This spread, which reaches more than one order of magnitude in some sub-regions, reflects the intrinsic variability of precipitating electron spectra both within a given perijove and from one perijove to another. 
Despite this variability, the mean distributions show an overall monotonic power-law decay over the full JADE+JEDI energy range, with globally comparable amplitudes across sub-regions. This behaviour is not shared by every individual spectrum entering these averages. As illustrated in Figure~\ref{fig:spectra_PJ3_Juno}, a fraction of the individual distributions depart from a strict power-law decay between $\sim 1$ and $\sim 100\,\text{keV}$, where a flattening or a plateau of the differential flux is occasionally observed before the decay resumes at higher energies, these departures being largely smoothed out by the averaging over many spectra. An energy cutoff is visible beyond $\sim 500\,\text{keV}$ in most sub-regions, although it is less pronounced in the northern ME and northern OE regions and is not always clearly identifiable in the individual spectra (Figures~\ref{fig:spectra_PJ3_Juno} and \ref{fig:mean_spectra_JADE_JEDI}).
This consistency in spectral shape across a large and variable sample of individual measurements is precisely what motivates a statistical approach: rather than focusing on a single, well-constrained event, whose amplitude and shape may not be representative of the broader population, averaging over many perijoves and auroral crossings allows us to isolate the spectral shape that is common to the precipitating electron population in each sub-region, which is the primary quantity of interest for the phenomenological distribution developed in this study.

The mean energies\footnote{The mean energy $\langle E \rangle$ of each precipitating electron energy flux distribution $f(E)$ is computed as $\langle E \rangle = \int E\,f(E)\,\mathrm{d}E / \int f(E)\,\mathrm{d}E$.} $\langle E \rangle$ of the mean distributions range from $\sim 21\,\text{keV}$ in the northern ME region to $\sim 119\,\text{keV}$ in the southern PE region across the auroral sub-regions. In the PE region, these values are consistent with the direct measurements of \citet{Allegrini2020}, who report mean energies above $100\,\text{keV}$ outside the narrow main oval crossing region. For the ME region, we do not have directly comparable measurements from \citet{Allegrini2020}, as their mean energies in this region are derived from individual, narrow main oval crossings rather than from a spatial average over the full ME sub-region as defined in this study.

Our mean energies also differ by a factor of 2 to 3 from the mean energies derived by \citet{Benmahi2024a} from spectral inversions of Juno/UVS observations at Jovian auroral altitudes, with the UVS-inverted values being systematically higher. Part of this discrepancy is expected, since JADE and JEDI measure the electron energy flux distribution at Juno's orbital altitude, far above the planet, whereas the UVS-inverted energies reflect the electron population at the much lower atmospheric altitudes where the UV emission is actually produced. Between these two altitudes, the precipitating electrons can be further accelerated by processes such as inertial Alfvén waves \citep{Hess2010, Hess2013, Saur2018}, whistler waves \citep{elliott_acceleration_2018}, and ion and electron inverted-V structures \citep{Mauk2017, mauk_diverse_2018, Clark2018}, as noted by \citet{Benmahi2024a}. Although these processes can in principle also decelerate particles, the higher mean energies inverted from UVS observations compared to our in-situ values suggest that acceleration is the dominant effect along the path to the atmosphere in the auroral regions considered here.

\subsection{Fitting the mean distributions in each auroral sub-region}
\label{fit_of_the_mean_distributions}

\begin{table}[h!]
\centering
\caption{MCMC fit parameters of the classical single-kappa distribution.}
\label{tab:mcmc_kappa}
\footnotesize
\renewcommand{\arraystretch}{1.4}
\begin{tabular}{llccc}
\toprule
Region & H & $\mathcal{F}_\mathrm{tot}$ [mW/m$^2$] & $\langle E \rangle$ [keV] & $\kappa$ \\
\midrule
\multirow{2}{*}{\rotatebox{90}{\small Polar}}
& N & \vpm{7.046}{0.850} & \vpm{463.19}{133.8} & \vpm{2.0138}{0.0111} \\[6pt]
& S & \vpm{6.114}{0.817} & \vpm{464.52}{144.5} & \vpm{2.0140}{0.0180} \\
\midrule
\addlinespace[8pt]
\multirow{2}{*}{\rotatebox{90}{\small \parbox{1.8cm}{\centering Main\\emission}}}
& N & \vpm{7.089}{0.439} & \vpm{153.94}{24.7} & \vpm{2.0117}{0.0022} \\[6pt]
& S & \vpm{9.649}{0.886} & \vpm{257.82}{60.7} & \vpm{2.0127}{0.0055} \\
\midrule
\addlinespace[8pt]
\multirow{2}{*}{\rotatebox{90}{\small Outer}}
& N & \vpm{12.117}{1.313} & \vpm{177.72}{39.7} & \vpm{2.0124}{0.0037} \\[6pt]
& S & \vpm{13.238}{1.371} & \vpm{187.44}{46.6} & \vpm{2.0129}{0.0048} \\
\bottomrule
\end{tabular}
\end{table}

We fit the mean precipitating electron energy flux distributions in the six auroral sub-regions using MCMC (Markov Chain Monte Carlo), testing successively the classical single-kappa distribution and the 4-kappa distribution introduced in this study. In both cases, the MCMC likelihood is computed by comparing the base-10 logarithm of the observed and modeled fluxes distributions, rather than the linear values. This choice avoids the fit being dominated by low-energy fluxes, which are systematically larger given the power-law nature of the spectra, and ensures that low- and high-energy parts of the spectrum are weighted comparably. Details on the sampler configuration, walkers, burn-in, and prior ranges are given in Appendix~\ref{appendix_mcmc}. Throughout this section, we report the total precipitating energy flux $\mathcal{F}_\mathrm{tot} = \int E\, f(E)\, \mathrm{d}E$ rather than the free amplitude parameter $Q_0$ itself (see Appendix \ref{model_transport}), since $\mathcal{F}_\mathrm{tot}$ is the physically meaningful quantity. As $f(E)$ includes $Q_0$, $\mathcal{F}_\mathrm{tot}$ scales linearly with it, and $Q_0$ can be recovered from the tabulated $\mathcal{F}_\mathrm{tot}$ values.

We first fit the parameters $\langle E \rangle$, $\mathcal{F}_\mathrm{tot}$, and $\kappa$ freely using the single-kappa distribution defined in Equation~\ref{eq:kappa_distribution} (Appendix~\ref{model_transport}). In this study, we leave $\kappa$ free to test without prior assumptions whether this distribution can reproduce the combined JADE+JEDI spectra over the full $100\,\text{eV}$--$1\,\text{MeV}$ range, which extends significantly further toward low energies than the JEDI-only range on which $\kappa = 2.5$ was established. The results are summarized in Table~\ref{tab:mcmc_kappa}, and the best fits are shown as dashed blue curves in Figure~\ref{fig:mean_spectra_JADE_JEDI}. Despite a broad exploration of the parameter space, the classical single-kappa distribution fails to reproduce the mean JADE+JEDI spectra over the full $100\,\text{eV}$--$1\,\text{MeV}$ range, particularly at low energies and at the energy cutoff observed beyond $\sim 500\,\text{keV}$. 
This failure is quantified by the systematic convergence of $\kappa$ to the lower bound of the prior range, $\kappa \in\, ]2.01, 40[$ (Appendix~\ref{appendix_mcmc}), with best-fit values of $\kappa \approx 2.01$ in all auroral sub-regions and both hemispheres (Table~\ref{tab:mcmc_kappa}), well below the $\kappa = 2.5$ value used in the literature. This lower bound is not an arbitrary choice, but is set by the mathematical structure of Equation~\ref{eq:kappa_distribution} itself, in which the terms $(\kappa-2)^2$ and $2E/(\kappa-2)$ produce a singularity at $\kappa = 2$, consistent with the divergence of the mean energy $\langle E \rangle = 2E_0\kappa/(\kappa-2)$ at this same value. Lowering the prior bound further would not yield convergence toward a well-defined interior value, but would simply let $\kappa$ approach this singularity more closely, without stabilizing. The MCMC algorithm therefore pushes $\kappa$ as close to the singularity as the prior allows in an attempt to harden the spectral slope, but still cannot simultaneously account for the low-energy behavior and the high-energy cutoff, confirming that the classical single-kappa distribution is fundamentally inadequate for describing JADE+JEDI spectral distributions over the $100\,\text{eV}$--$1\,\text{MeV}$ range.

Precipitating magnetospheric electrons in auroral regions are often described using simple phenomenological distributions or linear combinations of such distributions. \citet{Grodent2001} used such combinations in the Jovian context. For diffuse aurora, they use a kappa distribution combined with a double Maxwellian describing soft electrons, a superposition of three components. For discrete aurora, they use a triple Maxwellian whose first component describes hard high-energy electrons. Similarly, using combined JADE and JEDI data near Ganymede's orbit, \citet{Paranicas2021} found that electron spectra in Jupiter's radiation belt environment required an extended kappa distribution \citep{Hawkins1998}, a two-term functional form already more flexible than the classical single-kappa distribution, although in a different physical context than the auroral precipitation regions studied here. This shows that the morphological diversity of Jovian magnetospheric electron spectra generally cannot be captured by a single phenomenological distribution with a scalar parameter. We did not attempt to fit these alternative forms directly to the JADE+JEDI spectra in this study. The Maxwellian components used by \citet{Grodent2001} describe idealized electron populations rather than the broadband, non-thermalized distributions typically observed in the auroral regions considered here. The extended kappa distribution of \citet{Paranicas2021}, for its part, was established for a trapped radiation belt population near Ganymede's orbit rather than for a precipitating population within the auroral loss cone, a physical context that differs from the one studied here. In this study, this conclusion is reinforced by the failure of the classical single-kappa distribution demonstrated above, and extended to the $100\,\text{eV}$--$1\,\text{MeV}$ range now accessible through the JADE+JEDI combination. We therefore introduce a new, more flexible phenomenological distribution, the $N$-kappa distribution, defined as a linear combination of $N$ classical single-kappa distributions modulated by an exponential high-energy cutoff:
\begin{equation}
f_{N\kappa}(E) = \sum_{i=1}^{N} \omega_i\, f_{\kappa_i}(E)\, e^{-E/E_c},
\label{eq:kappa_modified}
\end{equation}
where $f_{\kappa_i}(E)$ is the classical single-kappa distribution defined in Equation~\ref{eq:kappa_distribution}, $\omega_i$ is the weight of the $i$-th component, $E_c$ is the cutoff energy controlling the high-energy behavior, and $N$ is the number of components. We set $N = 4$ in this study. This choice is motivated by the need for sufficient flexibility to reproduce the full diversity of individual JADE+JEDI spectra encountered across all auroral sub-regions, from the softest to the hardest, as further discussed below (Appendix~\ref{appendix_diversity_fits}). Fits performed with $N < 4$ components proved unable to reproduce this diversity satisfactorily across the full set of JADE+JEDI spectra (Appendix~\ref{appendix_diversity_fits}), which motivated retaining $N = 4$. The weights $\omega_i$ directly reflect the relative contribution of each component to the total precipitating energy flux. The mean energy of the distribution has an analytical expression given by:
\begin{equation}
\langle E \rangle = \sum_{i=1}^{N} 2\,\omega_i\,{\kappa_i}\, \langle E \rangle_i\, \frac{U(3,\, 3-{\kappa_i},\, \beta_i)}{U(2,\, 2-{\kappa_i},\, \beta_i)},
\label{eq:kappa_modified_mean_energy}
\end{equation}
where $\beta_i = {\kappa_i} \langle E \rangle_i / E_c$ and $U(a, b, z)$ is the confluent hypergeometric function of the second kind (Tricomi function).

The MCMC fit results obtained with the 4-kappa distribution are summarized in Table~\ref{tab:mcmc_kappa_modified}, the best fits are shown as green dashed curves in Figure~\ref{fig:mean_spectra_JADE_JEDI} and the prior ranges are given in Appendix~\ref{appendix_mcmc}. The 4-kappa distribution accurately reproduces the mean JADE+JEDI spectra in all six auroral sub-regions, with total energy fluxes $\mathcal{F}_\mathrm{tot}$ ranging from $\sim 9.5\,\text{mW/m}^2$ in the northern ME region to $\sim 28.7\,\text{mW/m}^2$ in the southern OE region. These values are broadly comparable to, though on the lower end of, the instantaneous fluxes reported by \citet{Clark2018}, who measured energy fluxes in the ME loss cone between $\sim 10$ and $\sim 100\,\text{mW/m}^2$, with a typical value of $\sim 50\,\text{mW/m}^2$, and by \citet{Allegrini2020}, who measured instantaneous fluxes typically between $\sim 10$ and $\sim 100\,\text{mW/m}^2$ during ME and PE crossings. This is expected, as the mean JADE+JEDI spectra used here result from averaging all distributions measured in each auroral sub-region over PJ3 to PJ35, which naturally smooths out both extreme flux events and local gaps, so that the resulting mean values are not directly comparable to instantaneous measurements.
The mean energies derived from the best-fit parameters using Equation~\ref{eq:kappa_modified_mean_energy} agree with the values computed directly from the mean spectra, ranging from $21.5 \pm 7.4\,\text{keV}$ in the northern ME region to $117.6 \pm 2.0\,\text{keV}$ in the southern PE region, against $\sim 21$ and $\sim 119\,\text{keV}$ respectively for the observed values.

The $\mathcal{F}_\mathrm{tot}$ values in the northern OE region ($24.900 \pm 0.328\,\text{mW/m}^2$) are markedly higher than in the northern PE ($14.354 \pm 0.462\,\text{mW/m}^2$) and northern ME ($9.537 \pm 0.16\,\text{mW/m}^2$) regions, although the OE value may be partly affected by the low-latitude contamination discussed in Section~\ref{JADE_JEDI_all_PJs}. Contrary to what might be expected from a simple hemispheric symmetry, the north-south comparison does not show a systematic trend across the three sub-regions. In the PE region, the northern and southern fluxes are comparable ($14.354 \pm 0.462$ vs $14.362 \pm 0.0887\,\text{mW/m}^2$), whereas the ME and OE regions both show higher fluxes in the southern hemisphere than in the northern hemisphere: ME ($9.537 \pm 0.16$ vs $16.653 \pm 0.367\,\text{mW/m}^2$) and OE ($24.900 \pm 0.328$ vs $28.686 \pm 0.479\,\text{mW/m}^2$). This north-south asymmetry, and its reversal relative to the PE region, is the subject of ongoing work and is beyond the scope of this study.

The cutoff energy $E_c$ is estimated beyond $\sim 3800\,\text{keV}$ in all six auroral sub-regions when fitting the mean spectra shown in Figure~\ref{fig:mean_spectra_JADE_JEDI} (Table~\ref{tab:mcmc_kappa_modified}), well above the $\sim 1\,\text{MeV}$ upper limit of the JADE+JEDI measurement range. As such, $E_c$ is not directly constrained by the mean spectra in any sub-region, and its value should be interpreted as an extrapolation rather than a direct measurement. This does not imply that $E_c$ is systematically unconstrained by the JADE+JEDI dataset as a whole. Fits performed on several hundred individual spectra across the auroral regions (Appendix~\ref{appendix_individual_fits}) show that $E_c$ falls below $1\,\text{MeV}$ in a fraction of cases, indicating that the mean spectra, being smoother than individual measurements, are less sensitive to the high-energy cutoff. Since the behavior of magnetospheric electrons beyond $1\,\text{MeV}$ is not documented in the literature, the extrapolated fit values reported in Table~\ref{tab:mcmc_kappa_modified} should be treated with caution.

The uncertainties on some parameters, in particular $\langle E \rangle_i$, $\omega_i$, and $E_c$, are sometimes large, especially in the ME and OE regions, reflecting partial degeneracy in the 4-kappa parameter space. This might suggest reducing the number of free parameters by fixing some of them. However, applying this distribution to fit several hundred individual spectra measured in the auroral regions, the distribution of best-fit values obtained independently for each parameter is broad and strongly non-unimodal (Appendix~\ref{appendix_individual_fits}). Also, as noted above, fits with fewer components in the $N$-kappa distribution failed to reproduce the diversity of JADE+JEDI spectra. This reflects the large morphological diversity of precipitating electron spectra observed by JADE+JEDI. Nevertheless, the total energy flux $\mathcal{F}_\mathrm{tot}$ remains well constrained in all sub-regions, and is the physically most significant parameter for electron transport modeling in Jupiter's atmosphere.

\begin{table*}[h!]
\centering
\caption{MCMC fit parameters of the 4-kappa distribution (Equation~\ref{eq:kappa_modified}) for the different Jovian auroral sub-regions.}
\label{tab:mcmc_kappa_modified}
\makebox[\textwidth][c]{%
\resizebox{1.04\textwidth}{!}{
\fontsize{10.5}{12.5}\selectfont
\renewcommand{\arraystretch}{1.4}
\begin{tabular}{llcccccccccccccc}
\toprule
\multirow{2}{*}{Region} & \multirow{2}{*}{H} & \multirow{2}{*}{$\mathcal{F}_\mathrm{tot}$ [mW/m$^2$]} & \multicolumn{3}{c}{$f_{\kappa_1}(E)$} & \multicolumn{3}{c}{$f_{\kappa_2}(E)$} & \multicolumn{3}{c}{$f_{\kappa_3}(E)$} & \multicolumn{3}{c}{$f_{\kappa_4}(E)$} & \multirow{2}{*}{$E_c$ [keV]} \\
\cmidrule(lr){4-6} \cmidrule(lr){7-9} \cmidrule(lr){10-12} \cmidrule(lr){13-15}
& & & $\langle E \rangle_{1}$ [keV] & $\kappa_1$ & $\omega_1$ & $\langle E \rangle_{2}$ [keV] & $\kappa_2$ & $\omega_2$ & $\langle E \rangle_{3}$ [keV] & $\kappa_3$ & $\omega_3$ & $\langle E \rangle_{4}$ [keV] & $\kappa_4$ & $\omega_4$ & \\
\midrule
\multirow{2}{*}{\rotatebox{90}{\small Polar}}
& N & \vpm{14.354}{0.462} & \vpm{0.0706}{0.00561} & \vpm{2.080}{0.0401} & \vpm{3.200\times10^{-6}}{1.15\times10^{-4}} & \vpm{4.332}{0.392} & \vpm{2.544}{0.378} & \vpm{5.656\times10^{-5}}{9.1\times10^{-4}} & \vpm{74.144}{12} & \vpm{2.981}{0.554} & \vpm{5.740\times10^{-4}}{0.0142} & \vpm{543.9}{135} & \vpm{2.567}{0.272} & \vpm{6.051\times10^{-5}}{0.00281} & \vpm{3815}{2160} \\[6pt]
& S & \vpm{14.362}{0.0887} & \vpm{0.0261}{0.00379} & \vpm{3.760}{0.976} & \vpm{8.182\times10^{-6}}{3.97\times10^{-5}} & \vpm{0.2572}{0.0157} & \vpm{2.372}{0.076} & \vpm{1.571\times10^{-4}}{7.51\times10^{-4}} & \vpm{4.506}{0.0889} & \vpm{2.119}{0.0226} & \vpm{0.01700}{0.0888} & \vpm{113.35}{2.99} & \vpm{7.381}{0.536} & \vpm{0.03337}{0.162} & \vpm{5832}{2230} \\
\midrule
\addlinespace[8pt]
\multirow{2}{*}{\rotatebox{90}{\small \parbox{1.8cm}{\centering Main\\emission}}}
& N & \vpm{9.537}{0.16} & \vpm{0.0200}{0.00966} & \vpm{3.665}{1.38} & \vpm{1.858\times10^{-6}}{2.58\times10^{-4}} & \vpm{0.2388}{0.0217} & \vpm{2.352}{0.116} & \vpm{8.159\times10^{-5}}{7.86\times10^{-3}} & \vpm{4.531}{0.39} & \vpm{2.612}{0.0914} & \vpm{3.843\times10^{-4}}{0.0257} & \vpm{83.680}{13.4} & \vpm{2.722}{0.943} & \vpm{0.00118}{0.0832} & \vpm{5961}{1980} \\[6pt]
& S & \vpm{16.653}{0.367} & \vpm{0.0839}{0.00472} & \vpm{2.096}{0.325} & \vpm{0.00160}{0.00168} & \vpm{3.035}{0.325} & \vpm{2.057}{0.0563} & \vpm{0.08497}{0.117} & \vpm{8.893}{50.8} & \vpm{8.026}{2.67} & \vpm{3.016\times10^{-6}}{0.0162} & \vpm{60.835}{10.6} & \vpm{3.498}{1.07} & \vpm{0.03607}{0.0897} & \vpm{8140}{1760} \\
\midrule
\addlinespace[8pt]
\multirow{2}{*}{\rotatebox{90}{\small Outer}}
& N & \vpm{24.900}{0.328} & \vpm{0.0478}{0.00292} & \vpm{2.055}{0.00943} & \vpm{0.00722}{0.012} & \vpm{4.009}{0.622} & \vpm{2.192}{0.711} & \vpm{0.05145}{0.0662} & \vpm{74.176}{6.22} & \vpm{2.588}{0.261} & \vpm{0.17968}{0.24} & \vpm{325.01}{286} & \vpm{5.543}{2.53} & \vpm{0.00177}{0.168} & \vpm{5758}{2160} \\[6pt]
& S & \vpm{28.686}{0.479} & \vpm{0.0934}{0.00555} & \vpm{2.141}{0.0504} & \vpm{0.02633}{0.00686} & \vpm{0.01048}{0.00115} & \vpm{4.360}{0.447} & \vpm{0.00140}{0.00029} & \vpm{4.594}{0.24} & \vpm{2.322}{0.0496} & \vpm{0.29205}{0.0886} & \vpm{114.10}{9.26} & \vpm{5.263}{1.57} & \vpm{0.78946}{0.0835} & \vpm{4055}{1910} \\
\bottomrule
\end{tabular}}}
\end{table*}

\subsection{Electron transport sensitivity study}
\label{comparaison_between_distributions}

\begin{figure}[h!]
    \centering
    \includegraphics[width=9cm, keepaspectratio]{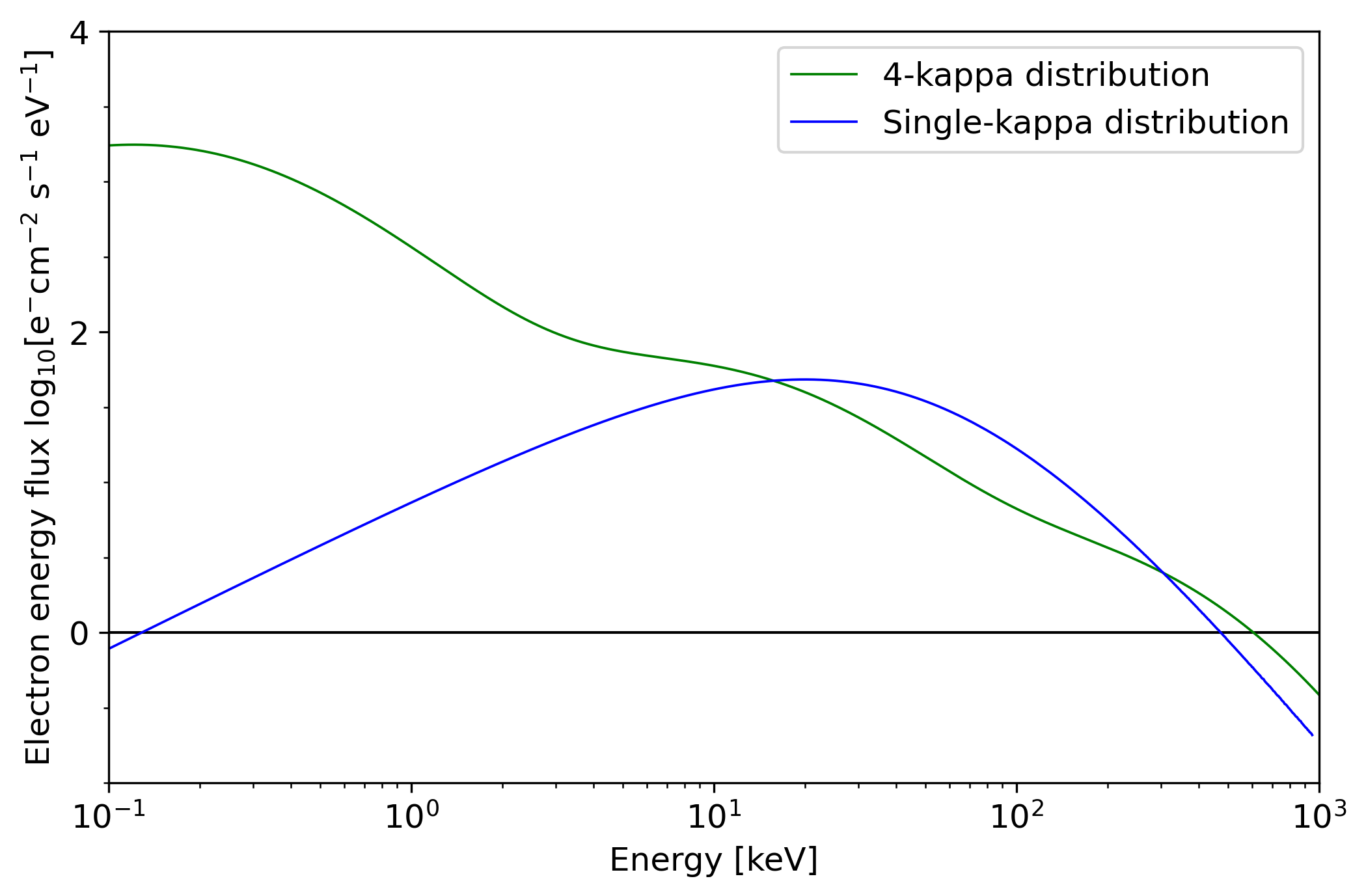}
    \caption{Comparison of the two energy flux distributions used as input to the TransPlanet transport code. The classical single-kappa distribution is shown in blue and the 4-kappa distribution in green. The y-axis is shown in log[$\text{cm}^{-2}\,\text{s}^{-1}\,\text{eV}^{-1}$]. For both distributions, the total energy flux is set to $\mathcal{F}_\mathrm{tot} = 1\,\text{mW/m}^2$ and the mean energy to $\langle E \rangle = 200\,\text{keV}$. The shape of the 4-kappa distribution ($\kappa_i$ and relative weights $\omega_i$) is taken from the best fit to the northern PE region (Table~\ref{tab:mcmc_kappa_modified}), with the mean energies $\langle E \rangle_i$ and the amplitude $Q_0$ rescaled so that $\langle E \rangle = 200\,\text{keV}$ and $\mathcal{F}_\mathrm{tot} = 1\,\text{mW/m}^2$.}
    \label{fig:transport_dist}
\end{figure}

\begin{figure}[h!]
    \centering
    \includegraphics[width=9cm, keepaspectratio]{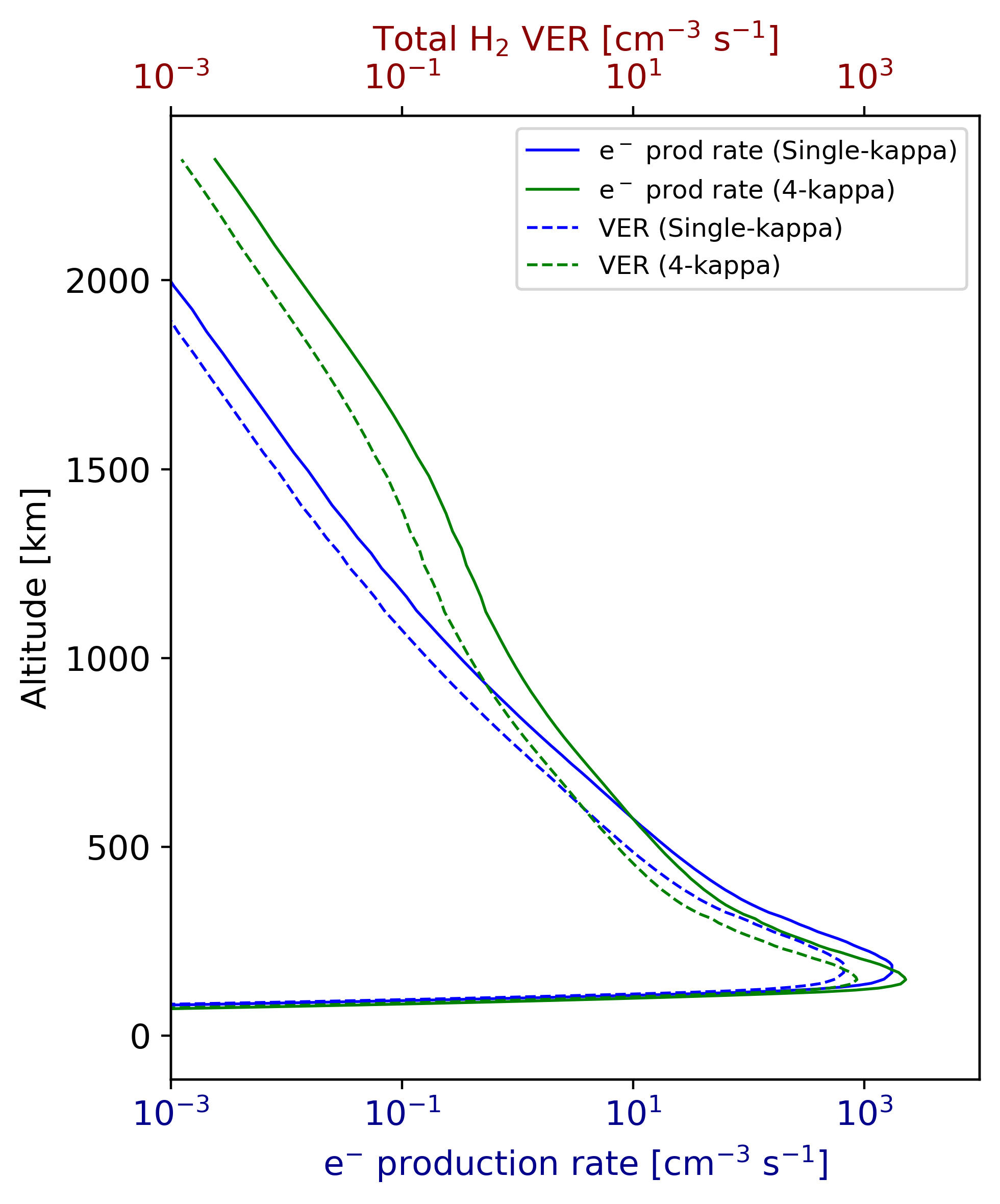}
    \caption{Vertical profiles of the electron production rate (solid curves, bottom axis) and of the total H$_2$ volume emission rate (VER, dashed curves, top axis) in Jupiter's atmosphere obtained with TransPlanet, for the classical single-kappa distribution (blue) and the 4-kappa distribution (green).}
    \label{fig:transport_eprod}
\end{figure}

To assess the impact of the shape of the precipitating electron energy flux distribution on atmospheric transport, we implement the 4-kappa distribution introduced in Section~\ref{results} in the TransPlanet code (Appendix~\ref{model_transport}) and compare the electron production rate and H$_2$ volume emission rate (VER) profiles obtained with both distributions. For both distributions, the total energy flux is set to $\mathcal{F}_\mathrm{tot} = 1\,\text{mW/m}^2$ and the mean energy to $\langle E \rangle = 200\,\text{keV}$, so that the two precipitations are comparable in terms of deposited energy. The shape of the 4-kappa distribution is taken directly from the best-fit parameters obtained for the northern PE region (Table~\ref{tab:mcmc_kappa_modified}), rescaled to match this common $\langle E \rangle$ and $\mathcal{F}_\mathrm{tot}$, so that the comparison reflects a spectral shape actually observed in the JADE+JEDI dataset rather than an arbitrary choice of parameters.

Figure~\ref{fig:transport_dist} shows the two input distributions. The 4-kappa distribution dominates at low energies ($\lesssim 15\,\text{keV}$) and in the high-energy tail ($\gtrsim 300\,\text{keV}$), while the classical single-kappa distribution dominates only in the intermediate range ($\sim 15$--$300\,\text{keV}$), with its linear low-energy rise placing its peak near $\sim 20\,\text{keV}$. This higher proportion of high-energy electrons in the 4-kappa distribution is consistent with the exponential cutoff energies inferred from our fits of the JADE+JEDI data, which range from $\sim 3800$ to $\sim 8100\,\text{keV}$ depending on the auroral sub-region (Table~\ref{tab:mcmc_kappa_modified}). The classical single-kappa distribution, which has no exponential cutoff, falls off faster than the 4-kappa distribution beyond $\sim 300\,\text{keV}$, whereas the 4-kappa distribution only drops sharply beyond $\sim 600\,\text{keV}$, reflecting its physically motivated high-energy cutoff.

Figure~\ref{fig:transport_eprod} shows the vertical profiles of the electron production rate and of the H$_2$ VER obtained for both distributions. As expected, the VER profile closely follows the electron production rate profile in both cases, since H$_2$ emission is directly proportional to the local electron production rate \citep{Benmahi2024a}. Both distributions produce qualitatively similar profiles, with a production rate that increases steadily from the top of the atmosphere down to a sharp low-altitude peak before dropping rapidly at the lowest altitudes. The two profiles differ mainly in the altitude of this peak and in the relative contribution of the higher altitudes. The single-kappa distribution peaks at $186.9\,\text{km}$, whereas the 4-kappa distribution peaks about $40\,\text{km}$ deeper, at $148.7\,\text{km}$. Both peaks lie well below the hydrocarbon homopause, and the 4-kappa distribution additionally deposits comparatively more energy at high altitudes ($\gtrsim 1000\,\text{km}$), reflecting its larger proportion of both low- and high-energy electrons relative to the single-kappa distribution.

Despite this difference in vertical structure, the column-integrated quantities are nearly identical between the two distributions. For the classical single-kappa distribution, the column electron production rate is $2.0559\times10^{10}\,\text{e}^-\,\text{cm}^{-2}\,\text{s}^{-1}$, and the column VER, i.e. the total H$_2$ brightness, is $8.10\,\text{kR}$. For the 4-kappa distribution, the column electron production rate is $1.9740\times10^{10}\,\text{e}^-\,\text{cm}^{-2}\,\text{s}^{-1}$, and the column VER is $7.72\,\text{kR}$. This near-equivalence is expected, since both distributions share the same total energy flux $\mathcal{F}_\mathrm{tot}$ and mean energy $\langle E \rangle$ by construction. The impact of the spectral shape is therefore not on the total amount of energy deposited or radiated, but on how this energy is vertically distributed in the atmosphere, with direct consequences for the local ionization and emission rates at a given altitude, as discussed below.

These results have direct implications for modeling Jupiter's UV auroral emissions and the impact of electron precipitation on the chemical composition and the thermal structure. Electron penetration below the CH$_4$ homopause ($\sim 300$--$400\,\text{km}$) increases the absorption of UV emissions in the $125$--$130\,\text{nm}$ band, which results in a higher UV color ratio \citep{Gustin2016, Gerard2019, Benmahi2024a}. Using the classical single-kappa distribution, which underestimates the proportion of high-energy electrons, therefore leads to a systematic underestimation of the simulated color ratio, and consequently to an overestimation of the mean energy of precipitating electrons when this energy is inverted from UV observations (e.g., \citealt{Benmahi2024a, Benmahi2024b, Vinesse2026}). This is consistent with \citet{Benmahi2024a}, who showed that using a monoenergetic distribution leads to an underestimation of the mean electron energy by a factor of 3--5 compared to a kappa distribution. Using the 4-kappa distribution should therefore provide an even more realistic estimate. The altitude at which the energy is deposited also determines where the ionization and dissociation of H$_2$ and CH$_4$ take place, and therefore the altitude at which auroral ion-neutral chemistry and the associated atmospheric heating are initiated, both of which control the composition and the thermal structure of the auroral upper atmosphere \citep{Sinclair2017, Sinclair2018}.
More generally, the difference in energy deposition profiles between the two distributions has implications for computing Pedersen and Hall ionospheric conductances in the auroral regions, which depend directly on the ion density profiles resulting from electron impact ionization \citep{Gerard2020, Gerard2021, Sicorello2025}. Recent studies have shown that using a more realistic broadband distribution significantly changes the computed conductances \citep{Sicorello2025}, and quantifying this impact for Jupiter's ionospheric conductances represents a natural extension of this work.

\section{Conclusions}
\label{conclusion}

We presented a statistical analysis of precipitating electron energy flux distributions in Jupiter's auroral regions, combining in-situ JADE and JEDI data from PJ3 to PJ35. This combination covers the full $100\,\text{eV}$--$1\,\text{MeV}$ energy range. After mapping the measurements into the Jovian SIII reference frame using the JRM33 magnetic field model, which accounts for the contribution of the current sheet \citep{Connerney2020, Connerney2022}, we built mean energy flux distributions in six auroral sub-regions: PE, ME, and OE regions, at both north and south poles.

While a single kappa distribution with $\kappa = 2.5$ initially appeared appropriate to fit JEDI-only data acquired over the ME region \citep{Salveter2022, Benmahi2024b}, we show here that it becomes inadequate to fit the full combined JADE+JEDI spectrum, including the high-energy drop-off beyond $\sim 500\,\text{keV}$. In our MCMC fit, $\kappa$ systematically converges to the lower bound of the prior ($\kappa \approx 2.012$--$2.014$) in all sub-regions, confirming this inadequacy. To address this, we introduced the 4-kappa distribution, a linear combination of four classical single-kappa distributions modulated by an exponential high-energy cutoff (Equation~\ref{eq:kappa_modified}). This distribution accurately fits the mean JADE+JEDI spectra in all six auroral sub-regions, and is, to our knowledge, the first analytical distribution directly fitted to combined JADE+JEDI data over a statistically significant set of perijoves.

The mean energies computed directly from the mean spectra range from $\sim 21\,\text{keV}$ in the northern ME region to $\sim 119\,\text{keV}$ in the southern PE region, consistent with the direct measurements of \citet{Allegrini2020} in the PE region, and are recovered by the 4-kappa fits within their uncertainties ($21.5 \pm 7.4$ and $117.6 \pm 2.0\,\text{keV}$, respectively). The total energy fluxes $\mathcal{F}_\mathrm{tot}$ range from $9.537 \pm 0.16\,\text{mW/m}^2$ in the northern ME region to $28.686 \pm 0.479\,\text{mW/m}^2$ in the southern OE region, broadly comparable to, though on the lower end of, the instantaneous fluxes reported by \citet{Clark2018} and \citet{Allegrini2020}.

Comparing electron transport in Jupiter's atmosphere between the two distributions reveals differences in the vertical structure of the energy deposition. At equal total energy flux and mean energy, both distributions produce a sharp low-altitude peak in the electron production rate, but the 4-kappa distribution peaks about $40\,\text{km}$ deeper than the classical single-kappa distribution, at $148.7\,\text{km}$ against $186.9\,\text{km}$, and deposits comparatively more energy at high altitudes. The column-integrated production rate and H$_2$ brightness remain nearly identical between the two cases, so the impact of the spectral shape lies in the vertical redistribution of the deposited energy rather than in its total amount. This difference has direct implications for modeling UV auroral emissions, particularly the UV color ratio which is sensitive to the altitude of precipitating electron energy deposition \citep{Benmahi2024a, Benmahi2024b}, for modeling the impact of electron precipitation on the atmospheric chemical composition and thermal structure, and for computing Pedersen and Hall ionospheric conductances (e.g., \citealt{Sicorello2025}).

The 4-kappa distribution established in this study provides a rigorous analytical framework for future electron transport and Jovian auroral emission modeling. Its implementation in the TransPlanet code opens the way to a reassessment of the mean energies of precipitating electrons inverted from Juno/UVS UV observations, based on a more realistic spectral description than the classical single-kappa distribution.

\begin{acknowledgements}
B. Benmahi and V. Hue acknowledge support from the French government under the France 2030 investment plan, as part of the Initiative d’Excellence d’Aix-Marseille Université – A*MIDEX AMX-22-CPJ-04. French co-authors acknowledge the support of CNES to the Juno and Juice missions. B.B. is a Research Associate of the Fonds de la Recherche Scientifique - FNRS. Authors acknowledge support from the JAFAR project (ANR-25-CE49-6683). The authors acknowledge the AMDA science analysis system provided by the Centre de Données de la Physique des Plasmas (CDPP) supported by CNRS, CNES, Observatoire de Paris and Université Paul Sabatier, Toulouse, for facilitating access to the PDS data used in this study.
\end{acknowledgements}

\bibliographystyle{aa} 
\bibliography{biblio.bib} 

\appendix

\section{Electron transport model and phenomenological energy distributions}
\label{model_transport}

We use the kinetic code TransPlanet to simulate electron precipitation in Jupiter's atmosphere. TransPlanet belongs to the Trans* family of codes \citep{Lilensten1989, Blelly1996} and was adapted to Jupiter in previous work \citep{Benmahi2024a, Benne2024, Benmahi2025}. The code solves multi-stream electron transport in interaction with the neutral atmospheric constituents. Its coupling with a H$_2$ UV auroral emission model is described in detail in \citet{Benmahi2024a}. The atmospheric profiles of abundance and temperature are taken from the 1D model of \citet{Grodent2001}, spanning altitudes from the tropopause (corresponding pressure $\sim 100\,\text{mbar}$) to the upper thermosphere (corresponding pressure $\sim 10^{-9}\,\text{mbar}$), and including the neutral species H, H$_2$, He, and CH$_4$. In this study, we consider only magnetospheric electron precipitation and do not account for the horizontal variability of chemical abundances across the auroral regions, consistent with the use of a 1D atmospheric model.

The input electron flux distribution $f(E)$ used in TransPlanet is normalized to the total precipitating energy flux $\mathcal{F}_\mathrm{tot}$ (in $\text{mW}\,\text{m}^{-2}$) using the amplitude parameter $Q_0$. Three phenomenological distributions have been commonly used in the literature. The monoenergetic distribution concentrates all energy flux at a characteristic energy $E_0$:
\begin{equation}
f(E) = Q_0\, \delta(E - E_0),
\end{equation}
where $\delta$ is the Dirac distribution. The Maxwellian distribution describes a thermalized electron population with mean energy $E_0$:
\begin{equation}
f(E) = Q_0\, \frac{E}{2 E_0^2} \exp\left(-\frac{E}{E_0}\right).
\end{equation}
The kappa distribution of \citet{Coumans2002}, hereafter referred to as the classical single-kappa distribution, with its power-law tail, was used to describe the broadband electron populations characteristic of Jupiter's auroral regions:
\begin{equation}
f_{\kappa}(E, \langle E \rangle) = Q_{0} \cdot \frac{4}{\pi} \cdot \frac{\kappa(\kappa - 1)}{(\kappa - 2)^2} \cdot \frac{E}{\langle E \rangle} \cdot \frac{\langle E \rangle^{\kappa - 1}}{\left( \dfrac{2E}{\kappa - 2} + \langle E \rangle \right)^{\kappa + 1}},
\label{eq:kappa_distribution}
\end{equation}
where $E$ is the electron energy, $\langle E \rangle$ is the mean energy, $\kappa$ is the suprathermal parameter controlling the slope of the high-energy tail, and $Q_0$ is the free amplitude parameter of the fit, from which the total precipitating energy flux $\mathcal{F}_\mathrm{tot}$ defined above is derived (Section~\ref{fit_of_the_mean_distributions}). The characteristic energy $E_0$, corresponding to the peak energy of the distribution, is related to the mean energy by:
\begin{equation}
\langle E \rangle = 2 E_{0} \cdot \frac{\kappa}{\kappa - 2}.
\end{equation}
In the limit $\kappa \rightarrow \infty$, the kappa distribution reduces to a Maxwellian distribution. This distribution was initially developed by \citet{Coumans2002} to model the proton population generating terrestrial FUV auroras, and has since been widely adopted in the Jovian community to represent broadband magnetospheric electron populations. It is used in studies of precipitating electron characterization, electron transport modeling, and UV auroral emission modeling at Jupiter \citep{Salveter2022, Benmahi2024a, Benmahi2024b, Sicorello2025, Vinesse2026}, as well as for Ganymede's auroral emission modeling \citep{Benmahi2025}. Based on the JEDI-only data study of \citet{Salveter2022}, restricted to the $30$--$1200\,\text{keV}$ range over the first 20 perijoves, \citet{Benmahi2024b} estimated statistically that $\kappa \approx 2.5$, a value commonly fixed in the studies cited above. However, the combined JADE+JEDI dataset used in this study extends this energy range down to $100\,\text{eV}$ and covers a larger set of perijoves (PJ3--PJ35), which, as shown in Section~\ref{fit_of_the_mean_distributions}, requires a more complex modeling approach than the classical single-kappa distribution.

\section{MCMC fitting procedure}
\label{appendix_mcmc}

The MCMC fits are performed using the affine invariant ensemble sampler emcee, exploring the parameter space with 125 walkers and 1900 iterations for both phenomenological distributions (classical single-kappa distribution, and 4-kappa distribution). Each chain is initialized with a narrow Gaussian draw around a physically reasonable starting point for the parameters. In both cases the likelihood is computed by comparing the base-10 logarithm of the observed and modeled fluxes. 

The prior distributions are uniform over the following parameters. For the classical single-kappa distribution, $Q_0 \in\, ]0, 10^{12}[$, $\langle E \rangle \in\, ]0, 6\times10^{7}[\,\text{keV}$ and $\kappa \in\, ]2.01, 40[$, where $Q_0$ is the free amplitude parameter of the fit, from which the total precipitating energy flux $\mathcal{F}_\mathrm{tot} = \int E\, f(E)\, \mathrm{d}E$ is derived. 

For the 4-kappa distribution, $Q_0 \in\, ]10^6, 10^{11}[$, $\langle E \rangle_1 \in\, ]0.001, 0.1[\,\text{keV}$, $\langle E \rangle_2 \in\, ]0.01, 10[\,\text{keV}$, $\langle E \rangle_3 \in\, ]0.3, 300[\,\text{keV}$, $\langle E \rangle_4 \in\, ]50, 1000[\,\text{keV}$, $\kappa_i \in\, ]2.05, 13[$, $\omega_i \in\, ]10^{-6}, 1[$ and $E_c \in\, ]10, 10^{4}[\,\text{keV}$. Both models converge after about one thousand iterations, which defines the burn-in period discarded before parameter estimation. Only the walkers whose final log-probability remains above the median minus 300 are retained for parameter estimation, which excludes poorly converged chains, and among these walkers only the last 50 iterations are kept for both distributions, in both cases well after convergence.

\section{Diversity of individual JADE+JEDI spectral shapes and flexibility of the 4-kappa distribution}
\label{appendix_diversity_fits}

\begin{figure}[h!]
\centering
\includegraphics[width=9cm, keepaspectratio]{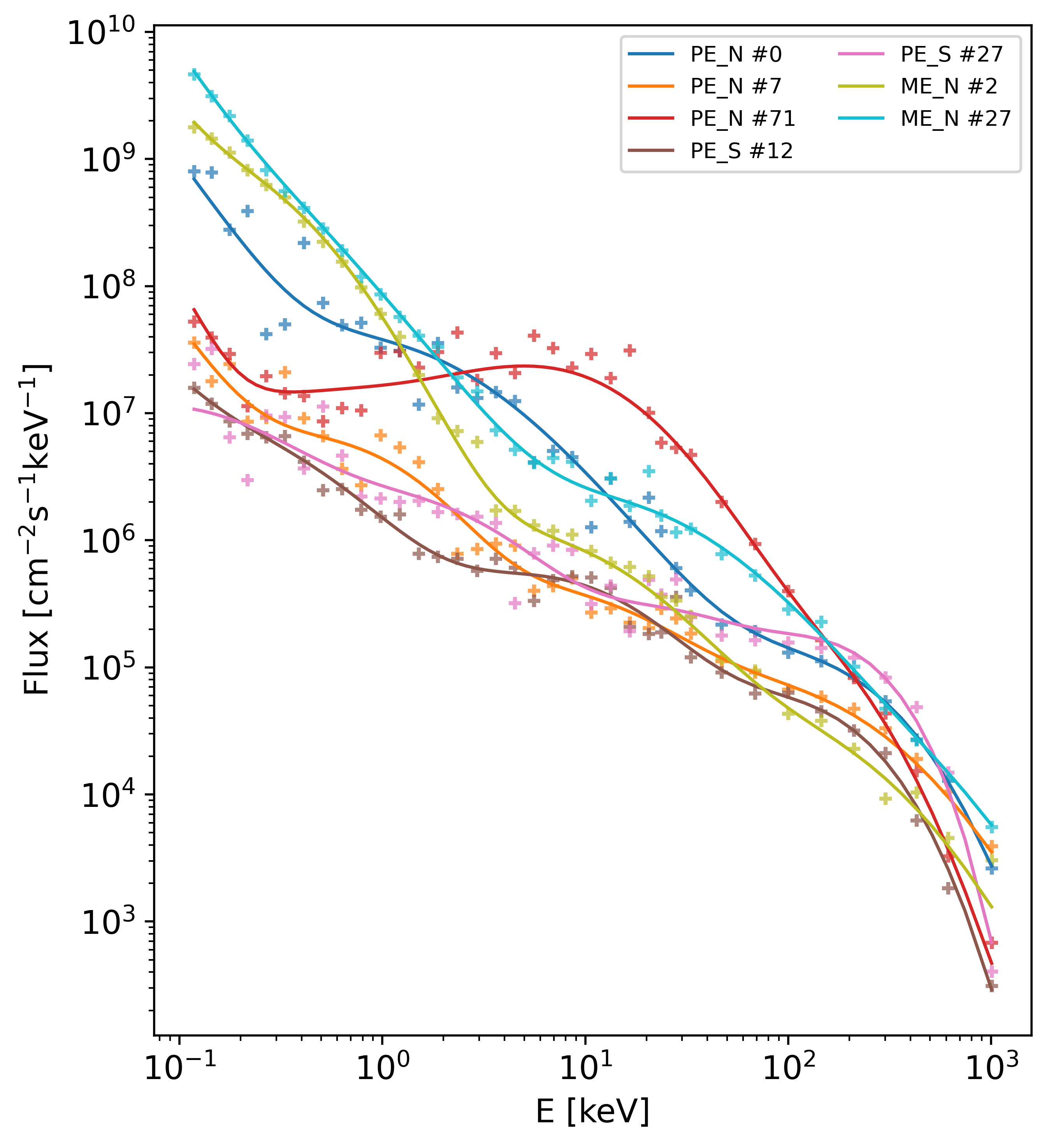}
\caption{4-kappa best fits (solid curves) to a sample of individual, randomly selected JADE+JEDI spectra (crosses) drawn from different auroral sub-regions and hemispheres, illustrating the diversity of spectral shapes encountered in the dataset and the flexibility of the 4-kappa distribution in reproducing them.}
\label{fig:diversity_fits}
\end{figure}

To illustrate the morphological diversity of individual JADE+JEDI spectra motivating the choice of $N = 4$ components in Equation~\ref{eq:kappa_modified} (Section~\ref{fit_of_the_mean_distributions}), Figure~\ref{fig:diversity_fits} shows the 4-kappa best fit obtained for a small sample of individual, randomly selected JADE+JEDI spectra drawn from different auroral sub-regions and hemispheres (PE\_N, PE\_S, and ME\_N). These spectra display markedly different shapes, ranging from a steep, strictly monotonic power-law decay (e.g., PE\_S \#12) to spectra with a pronounced bump or plateau structure at intermediate energies (e.g., PE\_N \#71), and spectra with a break in slope at low energy followed by a harder tail (e.g., ME\_N \#27). Despite this diversity, the 4-kappa distribution reproduces each individual spectrum closely over the full $100\,\text{eV}$--$1\,\text{MeV}$ range.

As noted in Section~\ref{fit_of_the_mean_distributions}, we also tested the $N$-kappa distribution with $N < 4$ components on this same set of individual spectra. These lower-order fits, not shown here, were unable to simultaneously reproduce the low-energy behavior, the intermediate-energy structures, and the high-energy cutoff visible in spectra such as those shown in Figure~\ref{fig:diversity_fits}, systematically leaving residual structure unaccounted for in at least part of the energy range. This confirms that a single or two-component kappa distribution is insufficiently flexible to capture the observed diversity of individual JADE+JEDI spectral shapes, and motivates retaining $N = 4$ components in this study.

\section{Statistical distribution of the 4-kappa best-fit parameters from individual spectra}
\label{appendix_individual_fits}

\begin{figure*}[h!]
\centering
\begin{subfigure}[b]{0.32\textwidth}
    \centering
    \includegraphics[width=\textwidth]{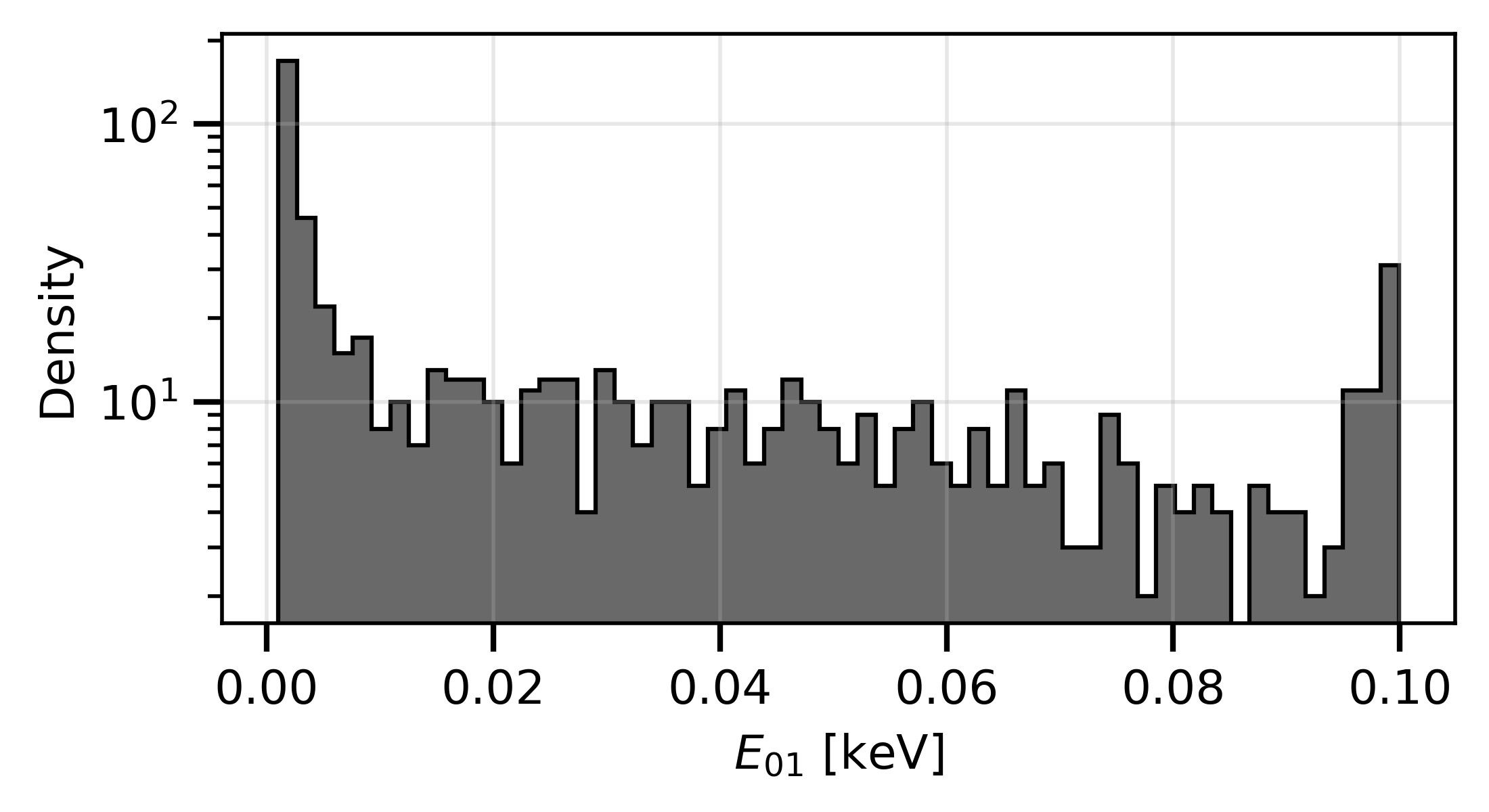}
\end{subfigure}
\hfill
\begin{subfigure}[b]{0.32\textwidth}
    \centering
    \includegraphics[width=\textwidth]{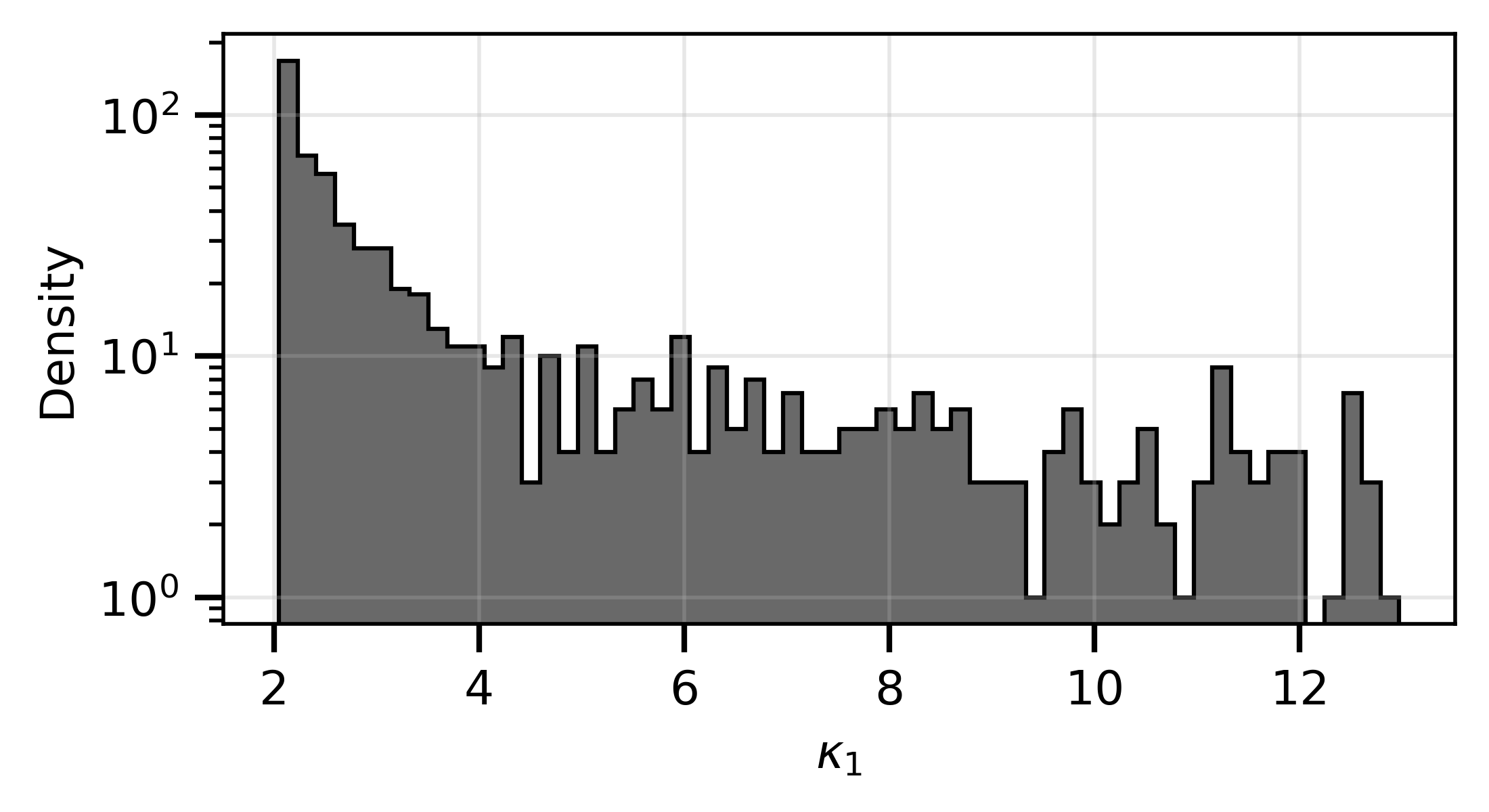}
\end{subfigure}
\hfill
\begin{subfigure}[b]{0.32\textwidth}
    \centering
    \includegraphics[width=\textwidth]{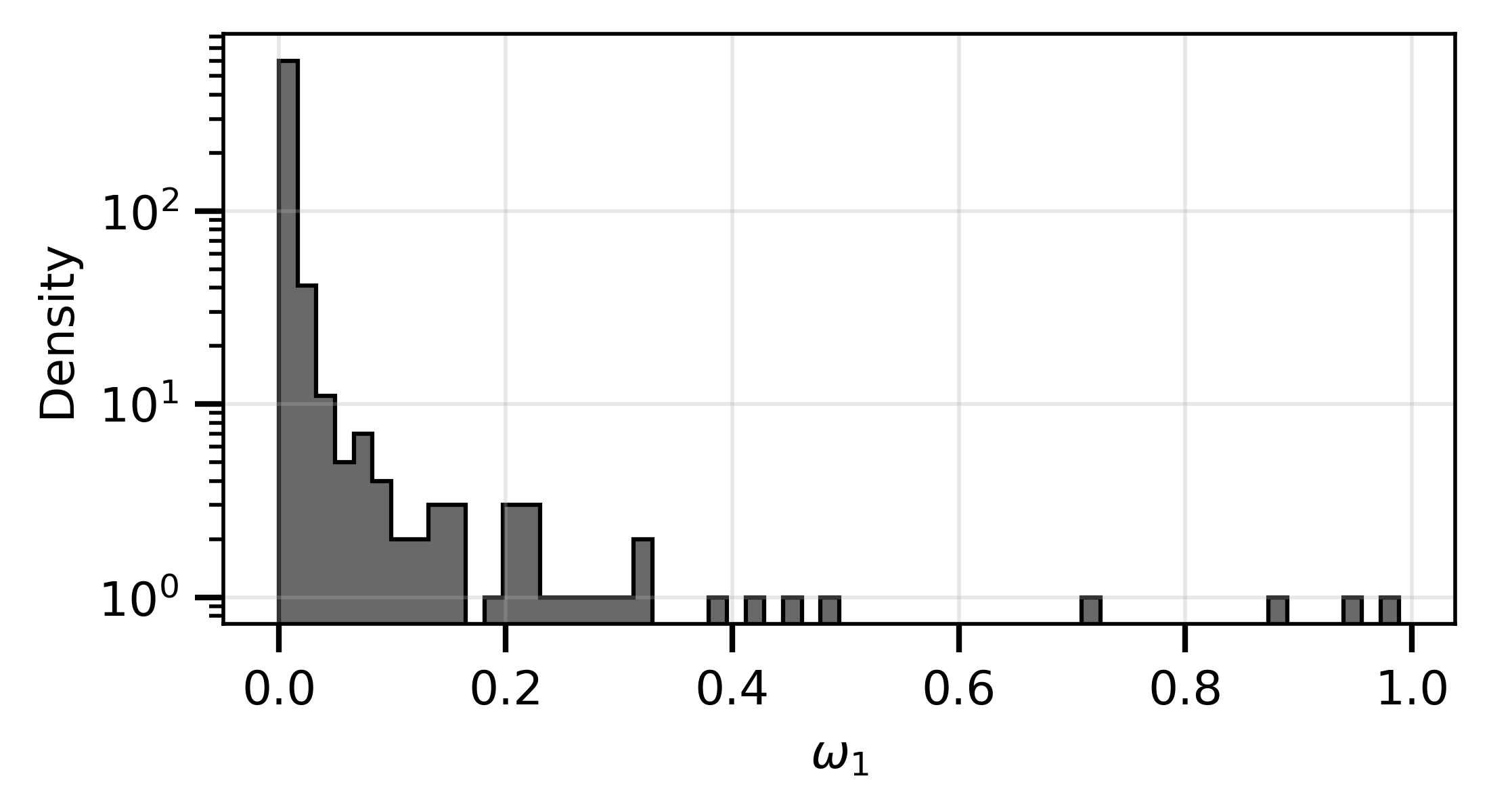}
\end{subfigure}

\begin{subfigure}[b]{0.32\textwidth}
    \centering
    \includegraphics[width=\textwidth]{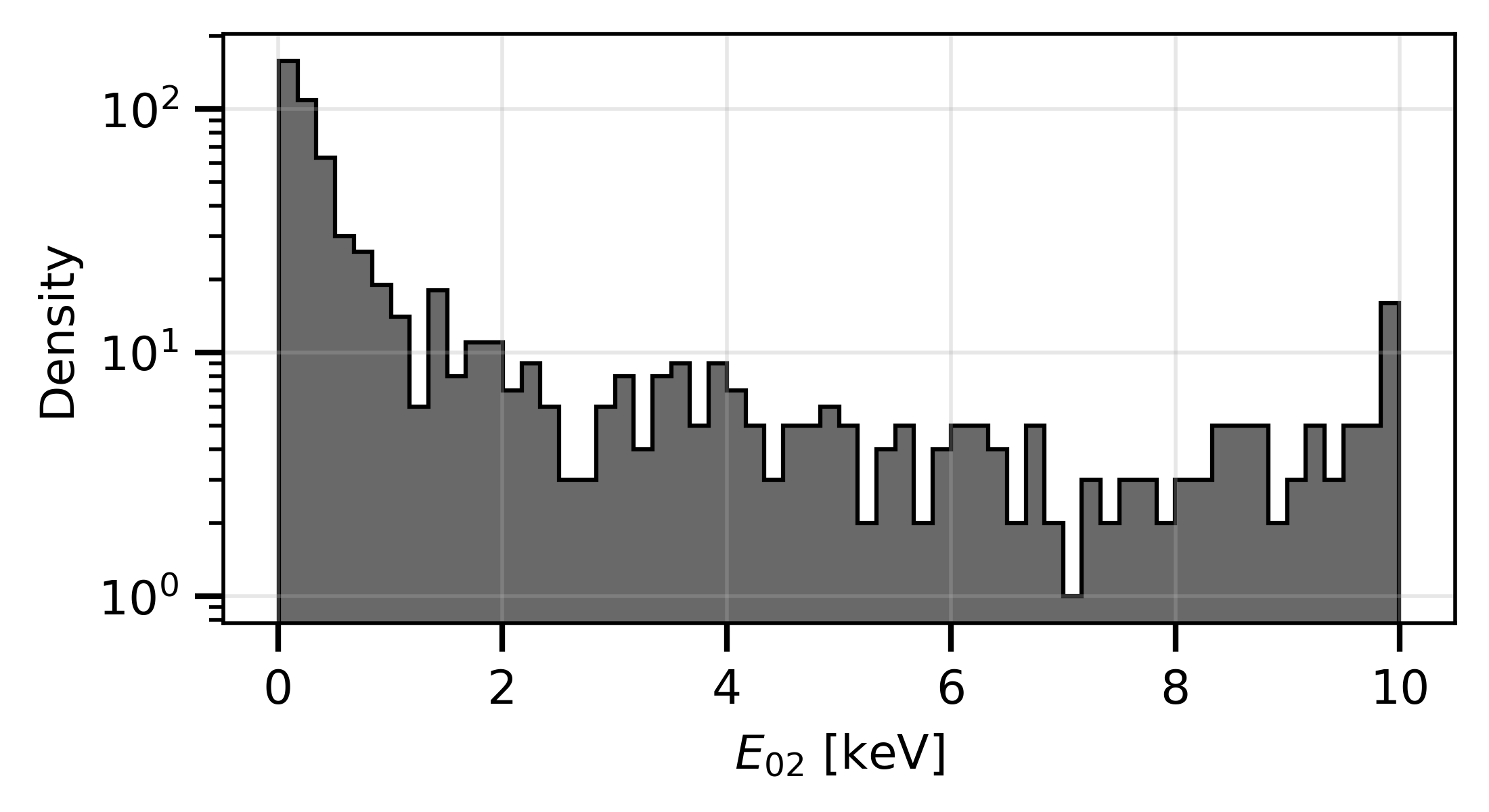}
\end{subfigure}
\hfill
\begin{subfigure}[b]{0.32\textwidth}
    \centering
    \includegraphics[width=\textwidth]{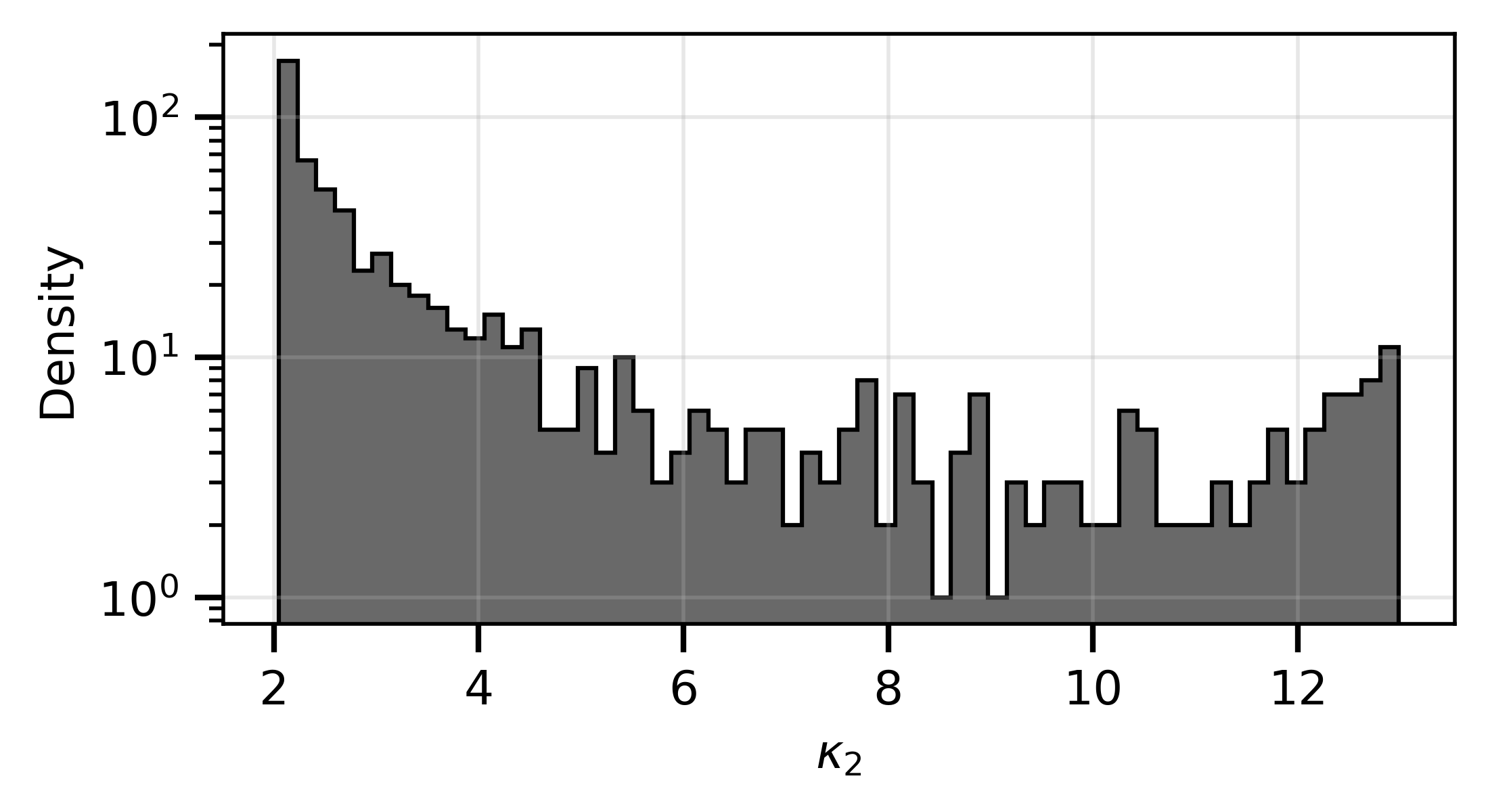}
\end{subfigure}
\hfill
\begin{subfigure}[b]{0.32\textwidth}
    \centering
    \includegraphics[width=\textwidth]{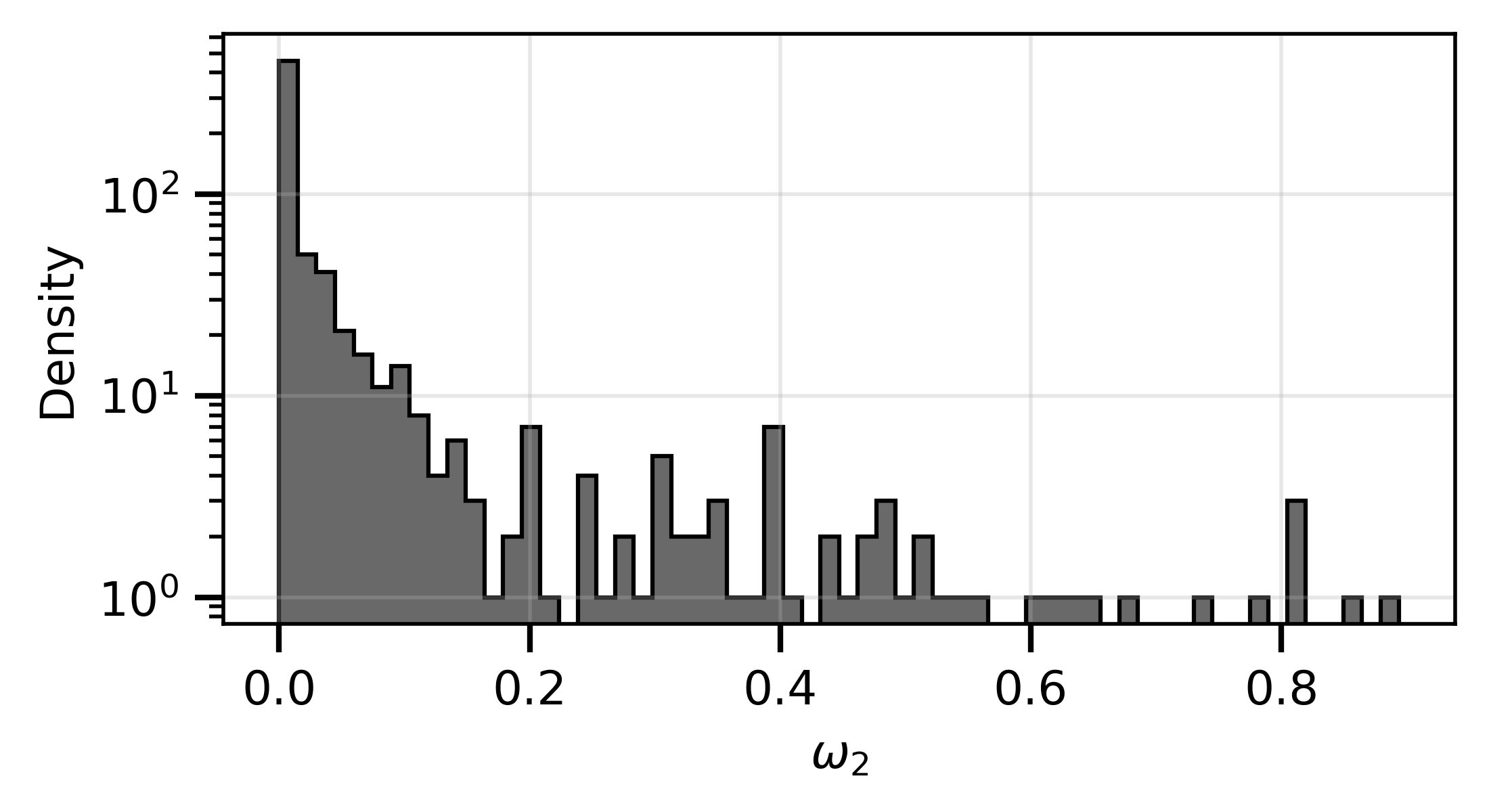}
\end{subfigure}

\begin{subfigure}[b]{0.32\textwidth}
    \centering
    \includegraphics[width=\textwidth]{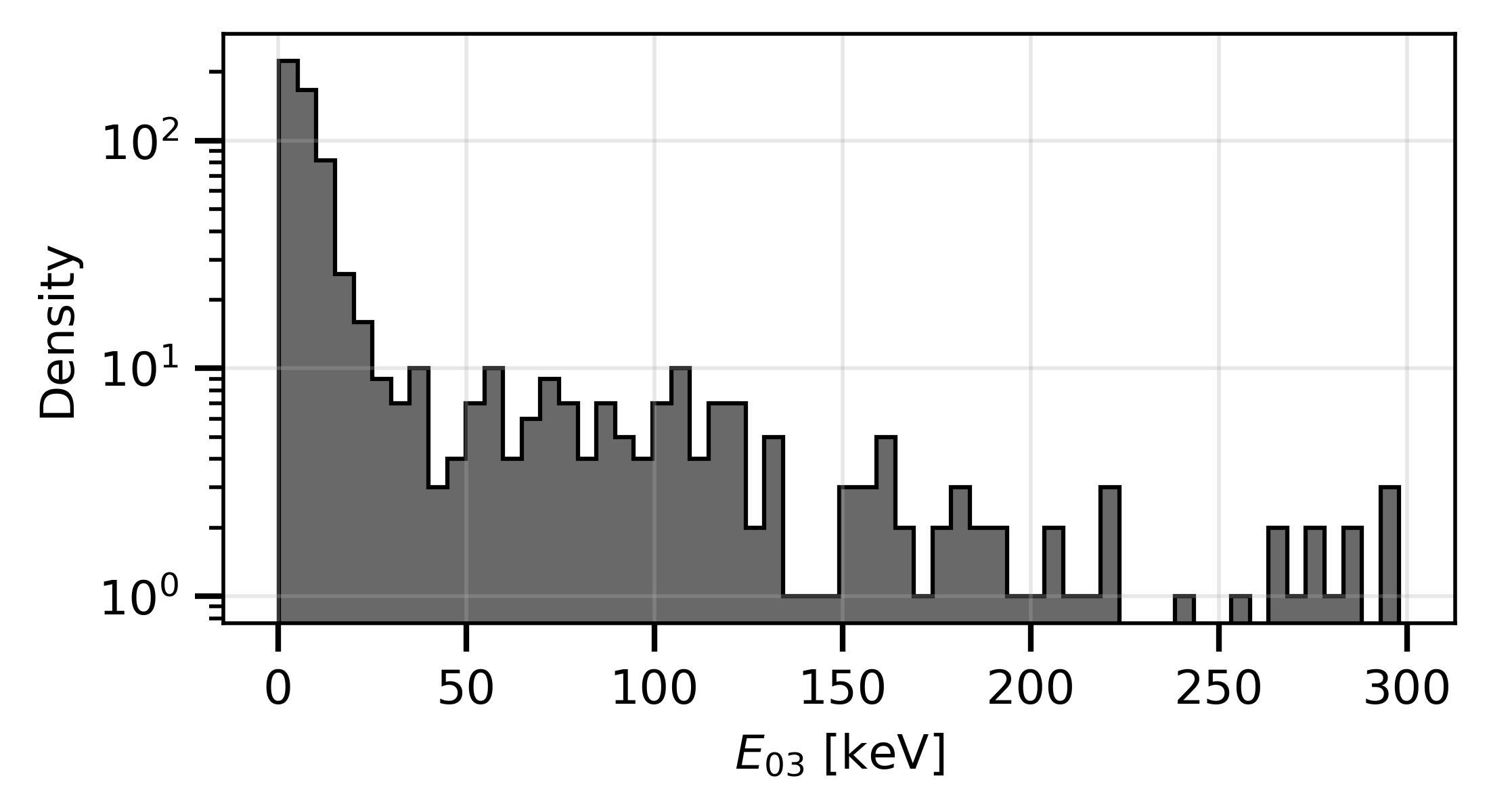}
\end{subfigure}
\hfill
\begin{subfigure}[b]{0.32\textwidth}
    \centering
    \includegraphics[width=\textwidth]{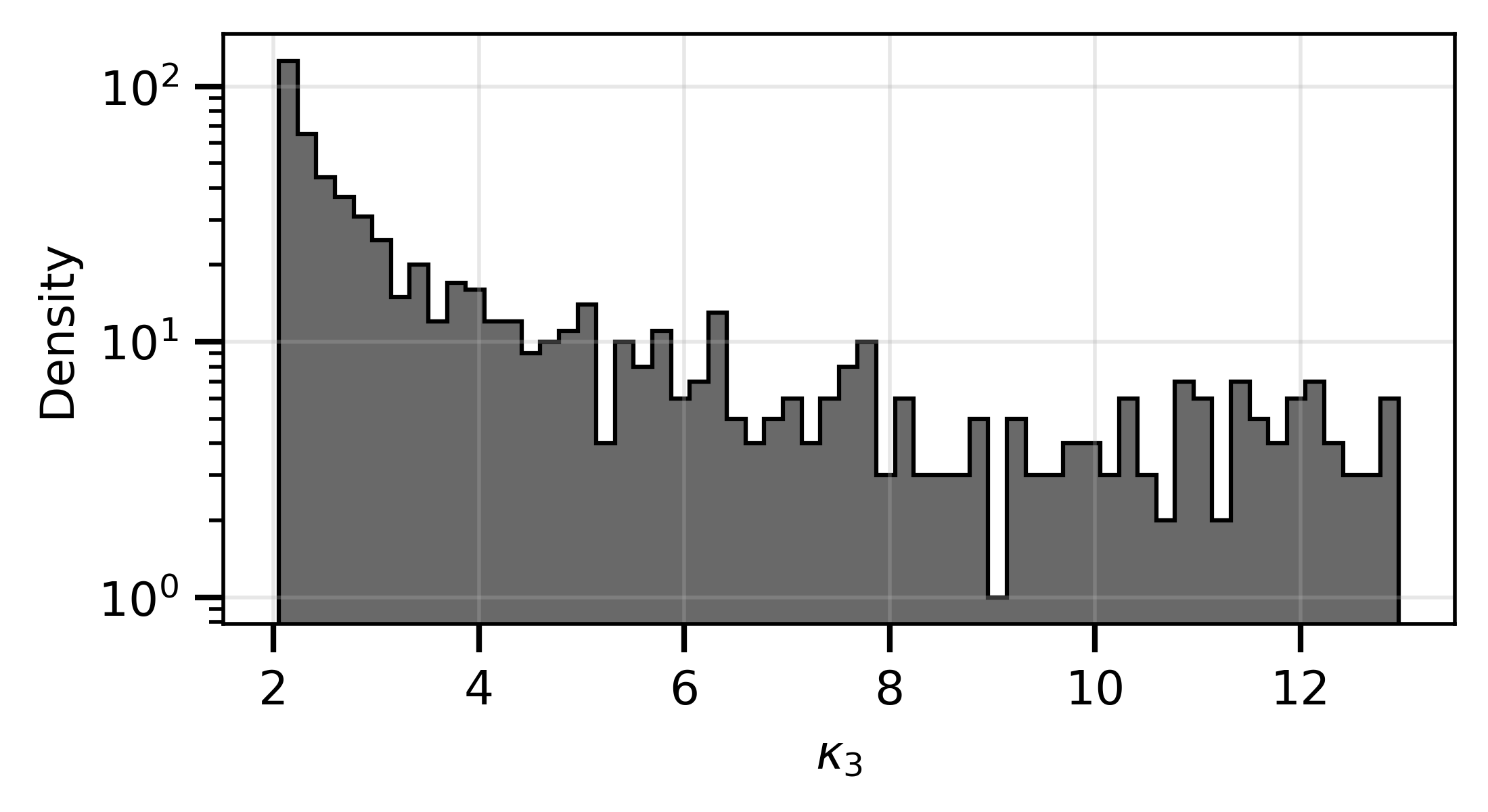}
\end{subfigure}
\hfill
\begin{subfigure}[b]{0.32\textwidth}
    \centering
    \includegraphics[width=\textwidth]{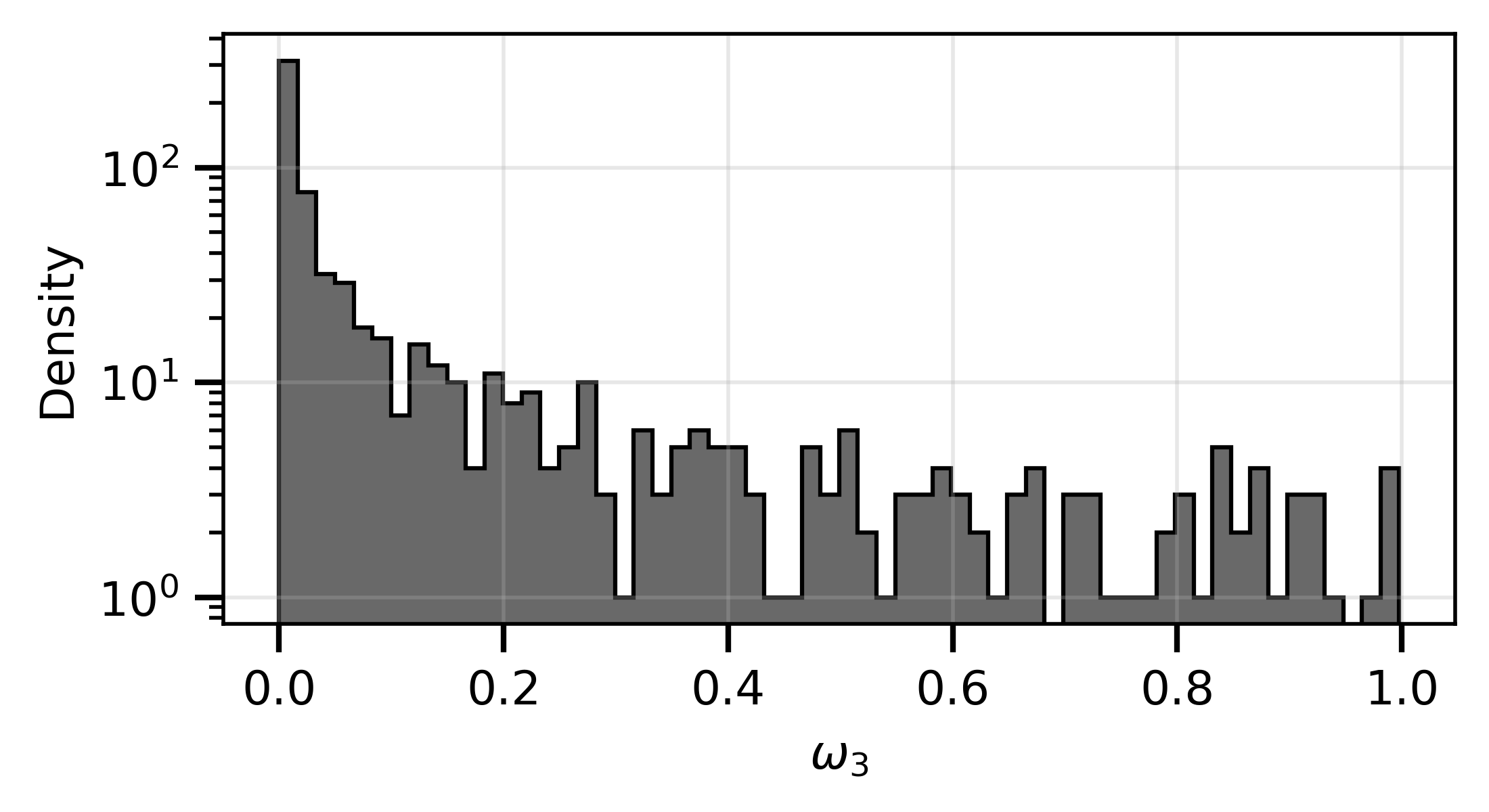}
\end{subfigure}

\begin{subfigure}[b]{0.32\textwidth}
    \centering
    \includegraphics[width=\textwidth]{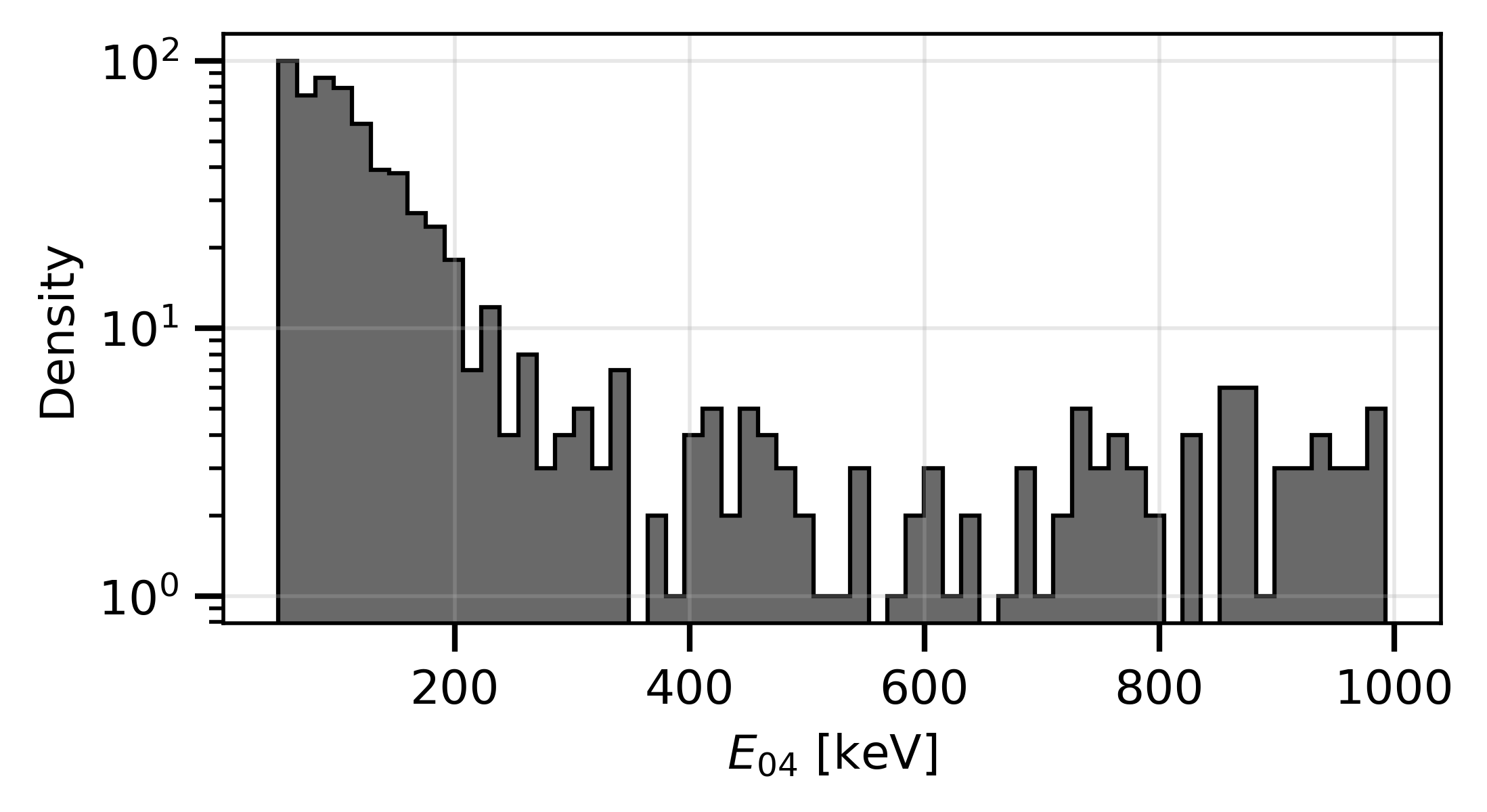}
\end{subfigure}
\hfill
\begin{subfigure}[b]{0.32\textwidth}
    \centering
    \includegraphics[width=\textwidth]{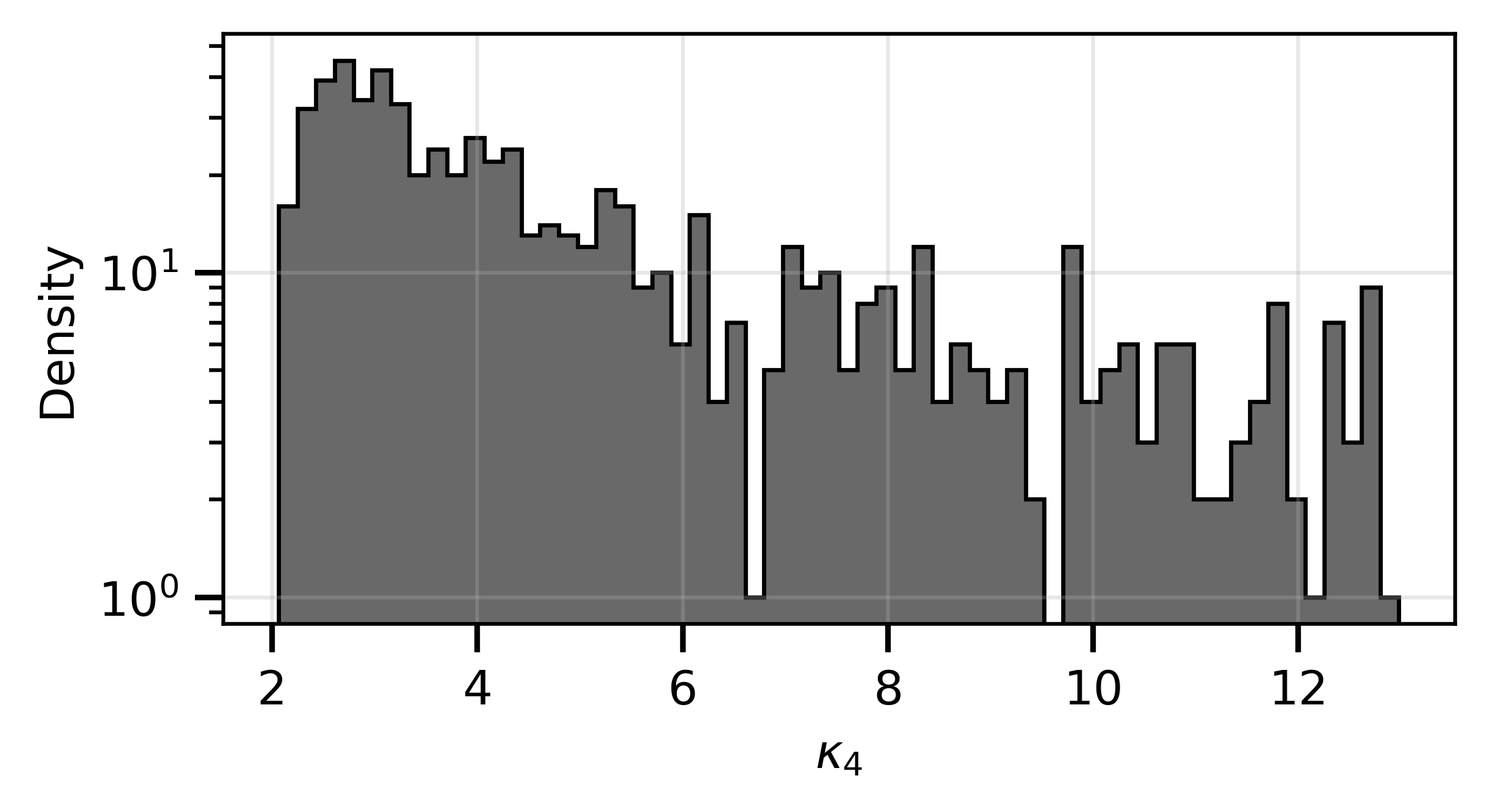}
\end{subfigure}
\hfill
\begin{subfigure}[b]{0.32\textwidth}
    \centering
    \includegraphics[width=\textwidth]{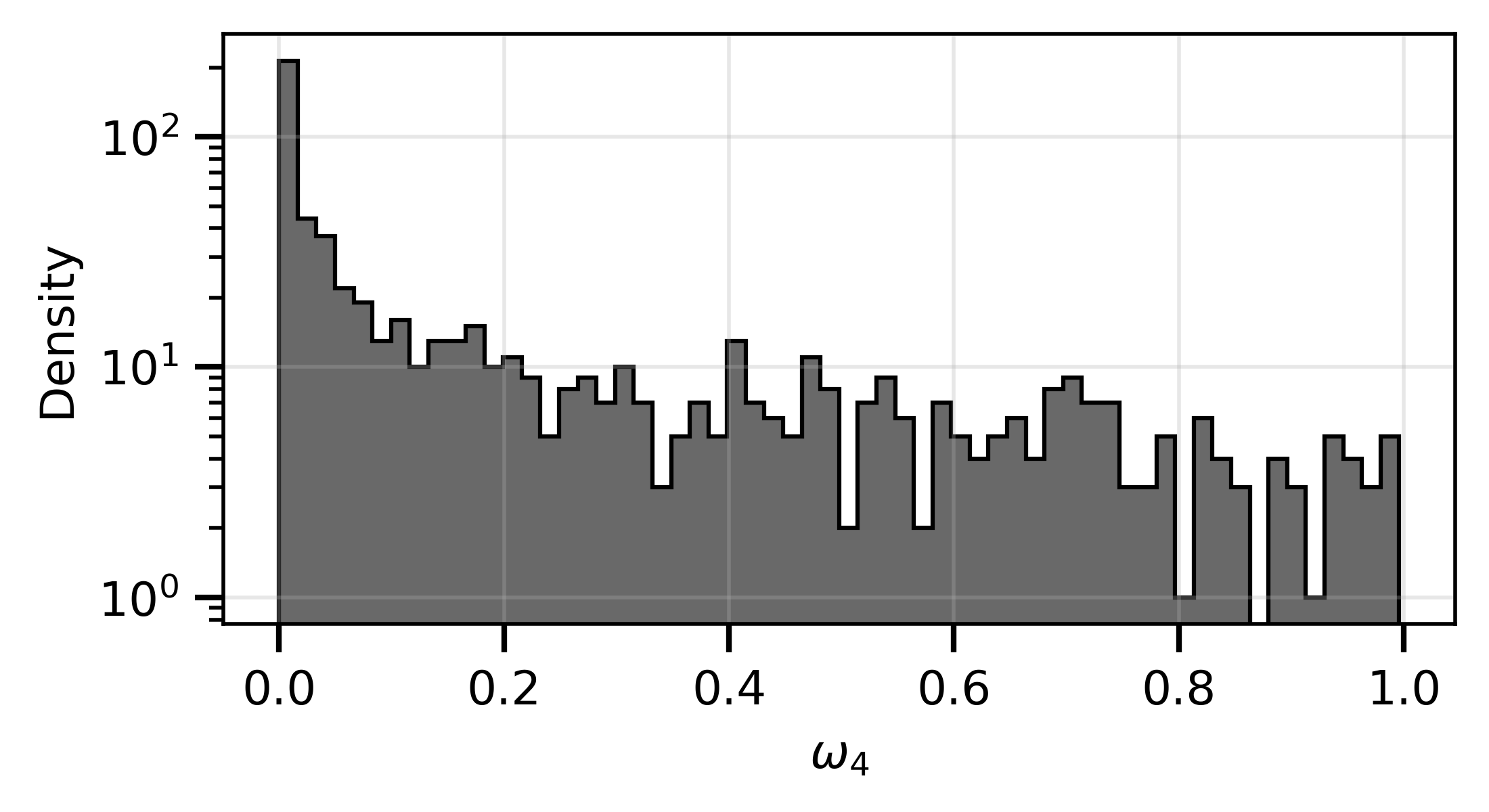}
\end{subfigure}

\begin{subfigure}[b]{0.32\textwidth}
    \centering
    \includegraphics[width=\textwidth]{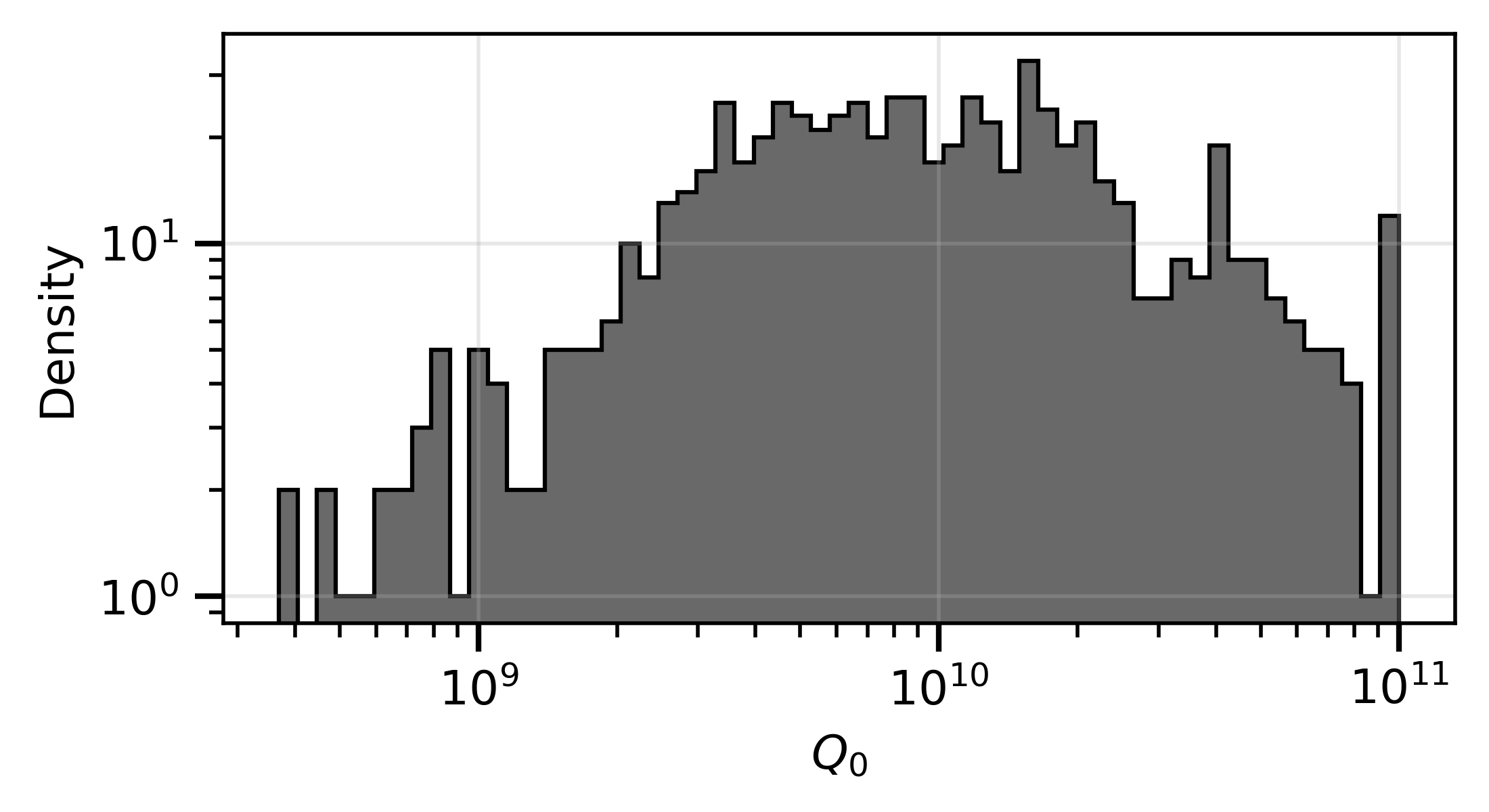}
\end{subfigure}
\hfill
\begin{subfigure}[b]{0.32\textwidth}
    \centering
    \includegraphics[width=\textwidth]{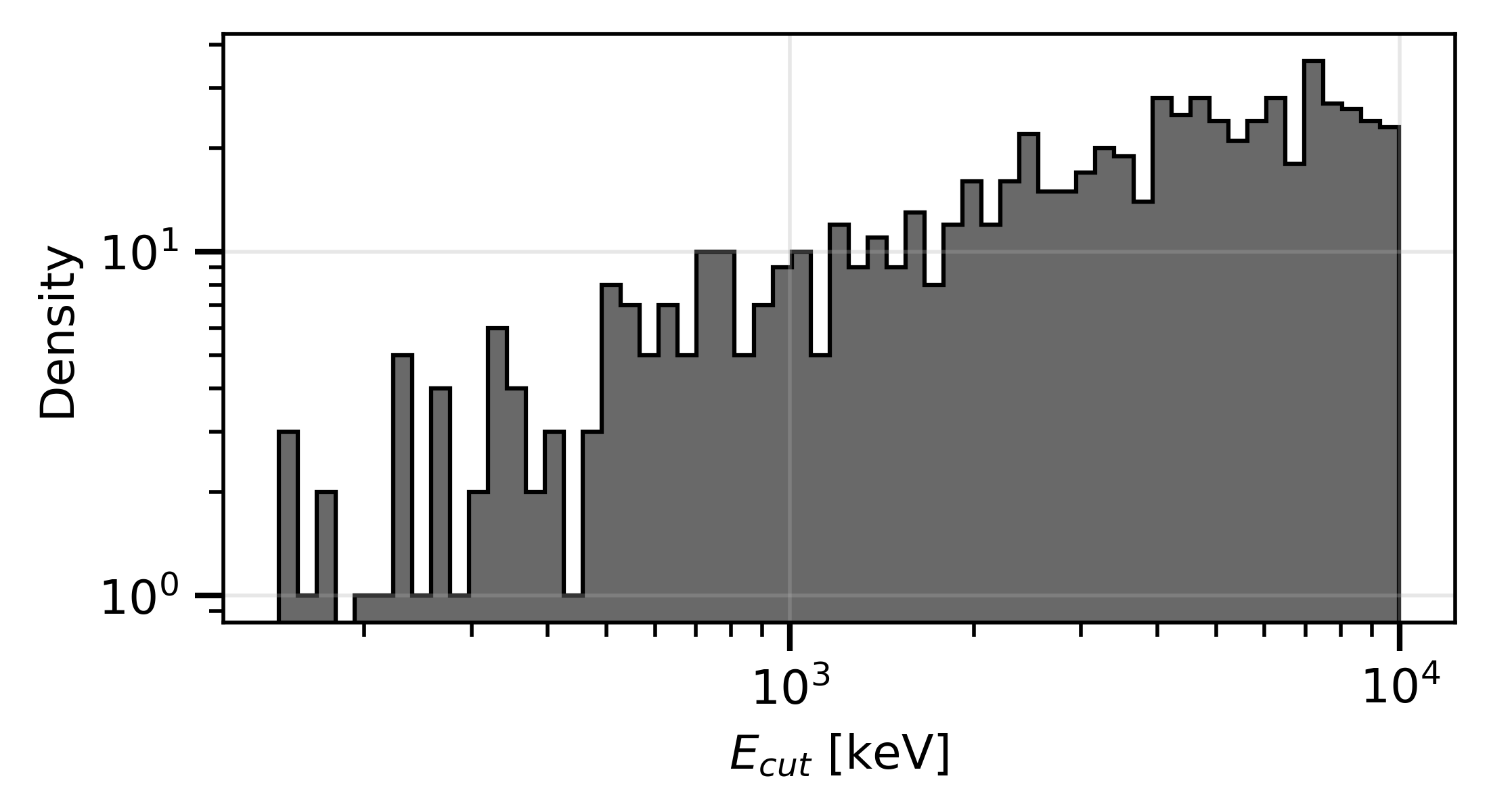}
\end{subfigure}

\caption{Probability density histograms (logarithmic vertical axis) of the best-fit parameter values of the 4-kappa distribution, obtained by independently fitting about 700 individual JADE+JEDI spectra across all six auroral sub-regions and both hemispheres. The first four rows show the mean energy $\langle E \rangle_i$, suprathermal index $\kappa_i$, and weight $\omega_i$ of each of the four kappa components ($i=1,\dots,4$), grouped by component. The bottom row shows the amplitude $Q_0$ and the cutoff energy $E_c$. Each histogram is built from the individual best-fit value obtained for each spectrum in the sample.}
\label{fig:individual_fits_params}
\end{figure*}

To further characterize the behavior of the 4-kappa distribution beyond the six mean spectra presented in Section~\ref{fit_of_the_mean_distributions}, we independently fit the 4-kappa distribution to about 700 individual, unaveraged JADE+JEDI spectra, sampled across all six auroral sub-regions in both hemispheres, using the same MCMC procedure and prior ranges described in Appendix~\ref{appendix_mcmc}. For each individual spectrum, we retain the best-fit value of each of the fourteen free parameters of the model, the amplitude $Q_0$, the four mean energies $\langle E \rangle_i$, the four suprathermal indices $\kappa_i$, the four weights $\omega_i$, and the cutoff energy $E_c$. Figure~\ref{fig:individual_fits_params} shows the resulting probability density histograms of these best-fit values across the full sample of individual spectra, with a logarithmic vertical axis to highlight the shape of each distribution over several orders of magnitude in occurrence.

None of the fourteen parameters shows a single, well-defined peak. Most distributions are broad, span most or all of their prior range, and show comparable probability density near both edges of the prior, without a dominant characteristic value. This is the case in particular for $\langle E \rangle_2$, $\langle E \rangle_3$, $\kappa_1$ to $\kappa_4$, and $\omega_1$ to $\omega_4$. This absence of a preferred value for any individual component parameter, across a sample of several hundred independently fitted spectra, is direct empirical evidence of the large morphological diversity of precipitating electron spectra observed by JADE+JEDI, and confirms that this diversity cannot be reduced to a small number of typical spectral shapes. This result also supports the choice of $N = 4$ components discussed in Section~\ref{fit_of_the_mean_distributions}, since a lower-dimensional model would need to average over this diversity rather than reproduce it. The cutoff energy $E_c$ shows an increasing density toward the upper edge of its prior range, consistent with the tendency, already noted in Section~\ref{fit_of_the_mean_distributions}, of $E_c$ to converge outside the JADE+JEDI measurement range. A non-negligible fraction of the individual fits nevertheless yields $E_c$ below $1\,\text{MeV}$, unlike the mean spectra for which $E_c$ systematically converges above this value.

\end{document}